\documentclass[proof]{WileyASNA-v1}

\articletype{Article Type}%

\received{September 2019}
\revised{October 2019}
\accepted{November 2019}

\newcommand{\degr}{\ensuremath{^\circ}}
\newcommand{\fdg}{\mbox{\ensuremath{.\!\!^\circ}}}
\newcommand{\arcmin}{\ensuremath{^\prime}}

\begin{document}

\title{A new approach to generate a catalogue of potential historical novae}

\author[1]{Susanne M Hoffmann}

\author[2]{Nikolaus Vogt}

\author[1]{Philipp Protte}

\authormark{Susanne M Hoffmann \textsc{et al}}

\address[1]{\orgdiv{Physikalisch-Astronomische Fakult\"at}, \orgname{Friedrich-Schiller-Universit\"at Jena}, \orgaddress{\country{Germany}}}

\address[2]{\orgdiv{Instituto de Física y Astronomía}, \orgname{Universidad de Valparaíso}, \orgaddress{\country{Chile}}}

\corres{Susanne M Hoffmann, FSU Jena. \email{susanne.hoffmann@uni-jena.de}}


\abstract{Ancient Chinese, Korean and Vietnamese observers left us records of celestial sightings, the so-called `guest stars' dated up to $\sim2500$ years ago. Their identification with modern observable targets could open interesting insights into the long-term behavior of astronomical objects, as shown by the successful identification of 8 galactic supernovae (SNe). Here we evaluate the possibility to identify ancient classical novae with presently known cataclysmic variables (CVs). For this purpose, we have developed a method which reconsiders in detail positions and sizes of ancient asterisms, in order to define areas on the sky that should be used for a search of modern counterparts. These areas range from a few to several hundred square degrees, depending on the details given in ancient texts; they should replace the single coordinate values given by previous authors.  Any appropriate target (CVs, X-ray binaries etc.) within these areas can be considered as a valid candidate for identification with the corresponding ancient event. Based on the original descriptions of several hundred old events, we selected those without movement and without a tail (to exclude comets) and which did not only visible within a certain hour (to exclude meteors). This way, we present a shortlist of 24 most promising events which could refer to classical nova eruptions. Our method is checked by applying it to the known SN identifications, leading to a margin of error between 0 and 4.5 degrees, meaning that some SN remnants lay exactly inside the areas given by the historical reports while in some other cases they are laying at considerable distances.}

\keywords{(stars:) novae, cataclysmic variables -- history and philosophy of astronomy}

\jnlcitation{\cname{%
\author{Susanne M Hoffmann}, 
\author{Nikolaus Vogt}, and 
\author{Philipp Protte}} (\cyear{2019}), 
\ctitle{A new approach to generate a catalogue of potential historical novae}, \cjournal{A.N.}, \cvol{2019; : }.}


\maketitle


\section{Introduction}\label{sec1}

\begin{figure}
	\includegraphics[width=\columnwidth]{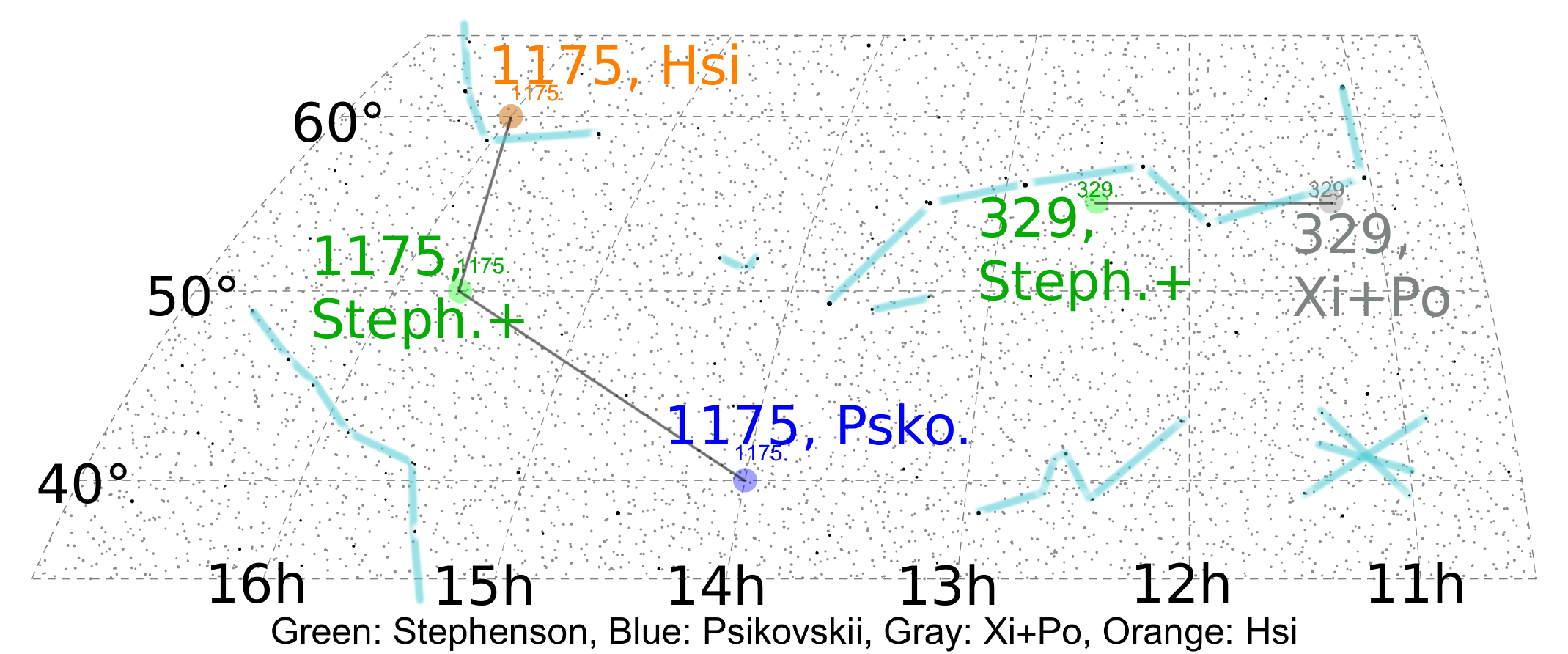}
    \caption{Coordinates of historical novae according to the lists of \citet{hsi}, \citet{xi+po}, \citet{pskovskii}, and \citet{stephenson} for two rather extreme cases: Four authors evaluating the same historical texts to derive positions of the phenomenon obtain results which differ by some 10 degrees. How can that be?}
    \label{fig:litVglN}
\end{figure}

\begin{figure}
	\includegraphics[width=\columnwidth]{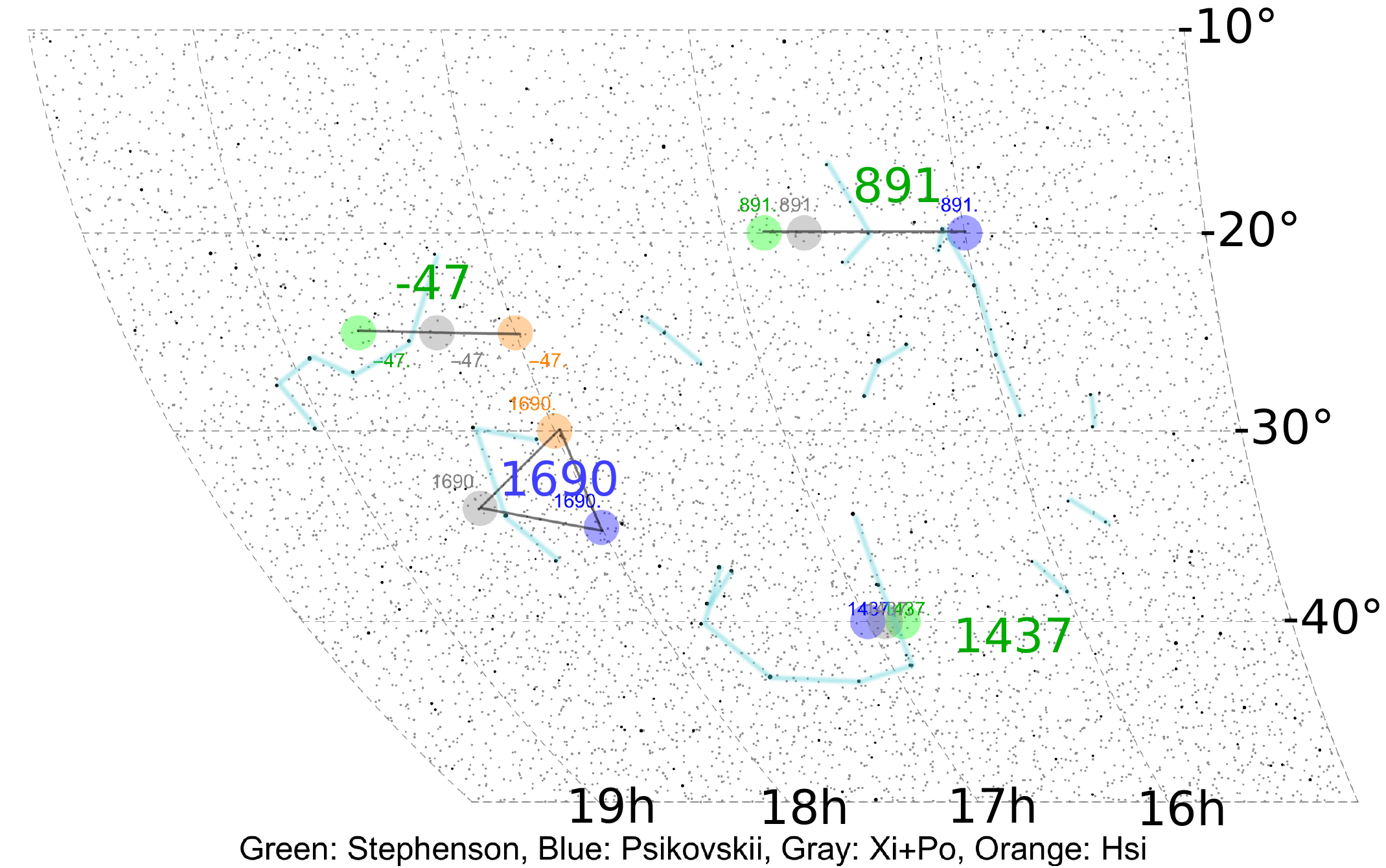}
    \caption{Additional to Fig.~\ref{fig:litVglN} another excerpt of the map in Fig.~\ref{fig:overview} showing the different localization of historical events by different recent authors. They give point coordinates (centres of the circles) but very likely considered some (unwritten) error ellipses around them which are indicated by the areas of the circle.}
    \label{fig:litvgl}
\end{figure}

During the last decades, some scholars presented nova catalogues extracted from Far Eastern guest star lists. Some of these catalogues contain coordinates to locate the historical guest stars (comparing the lists by \citet{hsi}, \citet{xi+po}, \citet{pskovskii}, \citet{stephenson}, and \citet{steph77}). Their results differ significantly from one list to the other (examples given in Fig.~\ref{fig:litVglN} and Fig.~\ref{fig:litvgl} do not show any pattern or systematic shift). These maps visualize large differences in identifying the locations of historical events, raising questions concerns the methods and accuracy of previous studies. In simple terms: Who is right? 

 In this study we will, therefore, present our own analysis of the historical records and give a less misleading base for further studies in this field of research. 

 Why is this topic worthy of investigation? Why do we need another catalogue of historical novae and why should that be more reliable than all the others which differ so much concerning their interpretation? During the last ten years, several scholars have used historical data to draw conclusions on astrophysical questions, e.\,g. concerning the development of close binary systems. For historical supernovae, this has been achieved for several decades (starting with \citet{schlier} and \citet{baade}), and for solar activity as well as for improving orbital elements of comets by usage of historical data e.\,g. \citet{NNC} and \cite{neuhaeuserKomet} also proved this concept to contribute to the current state of research. Thus, the idea of using old data caused the interest in research on cataclysmic variables culminating in papers by Shara$+$ in the recent years (e.\,g. \citet{shara2007}, \citet{shara2017_ATcnc_steph}, \citet{shara2017_nov1437}).
 
 \subsection{How many novae do we expect in 2000 years of history?}
  Considering an absolute brightness of novae of $(-7.5\pm0.5)$~mag and an apparent brightness of the nova of about 2~mag, the distance would be roughly 800~pc (neglecting extinction in the ISM) and therefore of the order of the thickness of the galactic disk. A sphere with this radius covers a volume of $2.2\cdot10^{9}$~pc$^3$ and assuming a space density of CVs of $2\cdot10^{-5}$~pc$^{-3}$ (according to \cite{belloni2018}) we derive a number of $4\cdot10^{4}$ CVs in this sphere.\footnote{cf. \citet{vogt2019}.} However, for apparent magnitudes of 3 or fainter the distance of the nova would be much more than 500~pc (half the thickness of the Milky Way galaxy). That means, that the galactic space density cannot be applied in the whole volume but only in the segment of the sphere close to the galactic equator. Thus, for the other values in Tab.~\ref{tab:NRnov} we assume a height above the galactic equator of 500~pc and compute the volume of the equatorial segment of the sphere. 
\begin{figure*}
	\includegraphics[width=\textwidth]{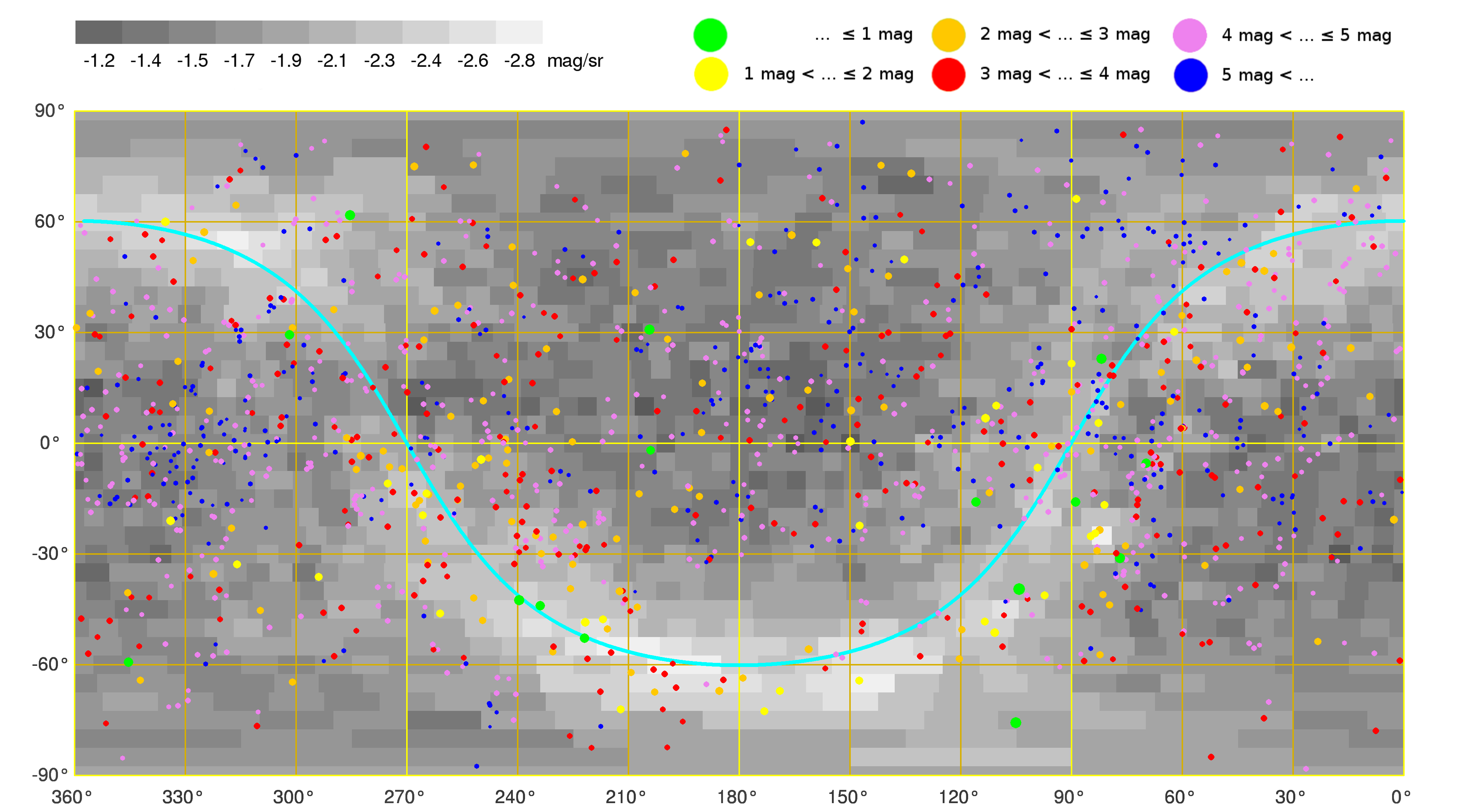} \\
	\includegraphics[width=\textwidth]{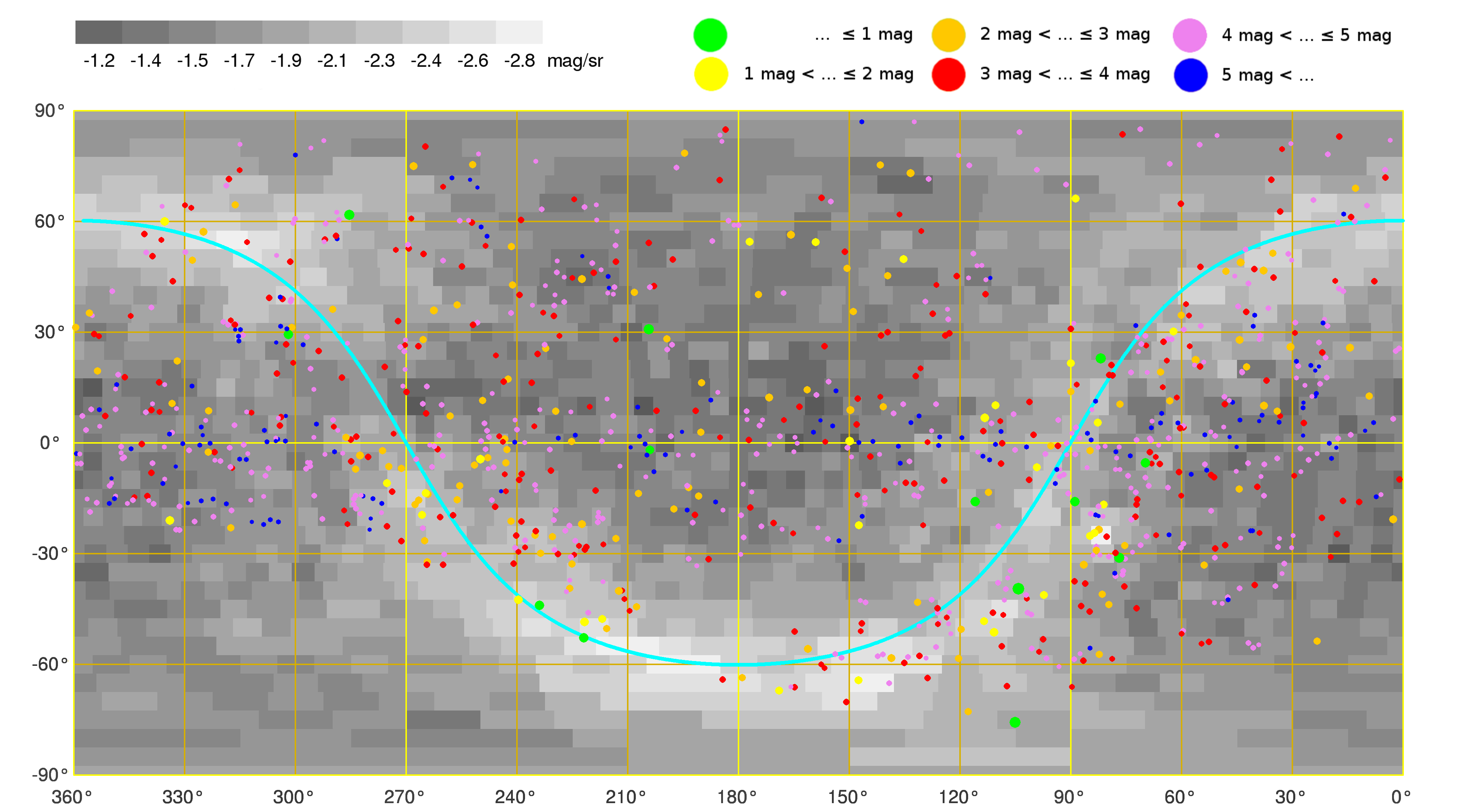}
    \caption{Distribution of stellar magnitudes in the ancient star catalogues of China (Suzhou: top) and Greece (Almagest: below). The colours of the stars indicate their visual magnitude (HIPPARCOS catalogue), the grey background tones display the general star density per steradian (the Milky Way is, therefore, visible in lighter shades and the Galactic equator indicated in cyan). These star catalogues are almost complete for brightnesses up to $4$~mag (cf. \citet[p.\,247]{smh2017} for the Greek one) but Chinese astronomers registered more faint stars ($\geq5$ mag in blue) than the Greek. We plotted the maps in ecliptical coordinates (equinox and epoch $+2000$) to make it more evident that the distribution of registered faint stars in both cases is more concentrated in the zodiac: The path of the Moon and planets certainly had an astrometrical function. In addition, both, the Greek and the Far Eastern sources reveal a lower abundance of faint stars towards the galactic plane.}
    \label{fig:distrMag}
\end{figure*}

\begin{table}
	\centering
	\caption{The total number of expected CVs ($\sharp$CVs) as function of the magnitude limit of  ancient observers (mag), distance of CVs (dist/pc) and space volume covered within the Milky way (Vol/pc$^3$). Details see text.}
	\label{tab:NRnov}
	\begin{tabular}{crrr} 
		\hline
		mag & dist/pc & vol/pc$^3$ & $\sharp$ CVs\\
		\hline
		2 & 800 & $2.2\cdot10^{9}$ & 42\,894\\
		3 & 1300 & $3.7\cdot10^{9}$ & 75\,927\\
		4 & 2000 & $16.0\cdot10^{9}$ & 319\,292\\
		5 & 3200 & $65.2\cdot10^{9}$ & 1\,303\,490\\
		\hline
	\end{tabular}
\end{table}

Depending on the rate of recurrence of nova outbursts in these systems (assumptions vary between $10^3$ to $10^6$ years with higher probability between $10^4$ to $10^5$) we expect a few ($\sim5$) to a few hundred ($\sim430$) novae per century. That is why we consider it worthwhile to look for nova observations in the corpus of astronomical reports among Far Eastern chronicle texts. This fact and the possible radii of the sphere in which we will find the candidates of historical novae we have already discussed in our literature study \citet{vogt2019} in more detail. Yet, we will not use any distance limitations as a selection criterion. Instead, we insist on phenomenological parameters only to identify our candidates -- such as the current brightness of the object and the known distribution of amplitudes of classical novae as displayed in Fig.~1 of our paper \citet{vogt2019}.  
 
 \subsection{On the celestial distribution of ancient guest star events}
 Standard assumptions on observations of classical novae assume a higher density close to the galactic plane and especially in the galactic centre implying a higher observational likelihood there (cf. \citet[p.\,1--2, especially Fig.~1.3]{bode}). A higher density of novae in an area of higher star density is indeed logical but this does not automatically imply more observations there: In an area with brighter background (because of milky way clouds) the naked eye observer will detect fainter novae less readily than in areas of only few bright stars. Thus, the probability of detecting a nova naked eye increases for high galactic latitudes because the detection limit is fainter and, thus, the radius of the space volume visible to the human observer is bigger. Hence, for high galactic latitudes, there are more stars in the spacial range and we expect more novae visible to the naked eye. 

 \textbf{Description of Fig.~\ref{fig:overview}}: To compare these expectations with the published lists of historical (ancient) novae (time span from 531~BCE to 1604~CE) we plotted all suggested coordinates by \cite{hsi,xi+po,pskovskii,stephenson,steph77} into an all-sky map (Fig.~\ref{fig:overview}). The frame of reference is equatorial, the Milky Way is visible as higher star density, and the ecliptic is also plotted (orange $-$sine). The map shows a higher number of historical detections at the galactic bulge and on the opposite side of the great circle of the galaxy (roughly in our constellation of Cassiopeia). Additionally, we discover a comparably high density along the ecliptic indicating the relevance of the zodiacal zone (the path of the moon) for the preservation of records (divinatory meaning). Even in the middle of the map, close to the galactic pole, there are some historical records.

 \textbf{Significance of the distribution of Fig.~\ref{fig:overview} in Chinese astral science}: The most crucial parts of the sky for diviners were the twenty-eight lunar lodges and the northern circumpolar region. The latter is the asterism enclosure called The Purple Forbidden Palace containing many smaller asterims like the Emperor, the Crown Prince, the Chief Judge, Royal Secretaries and Archives et cetera. The twenty-eight lunar lodges are RA-slides named after the asterisms which form the Four Holy Beasts; they corresponded to different polities in the known world in pre-imperial times, while the circumpolar region was associated with the central court in imperial times. Different parts of the sky mapped onto different seasons, regions, and days within the sexagenary calendar. Others were associated with legendary figures. The constellation Xuanyuan (in the middle of our map) bore an alternate name for the mythological Yellow Emperor Huangdi. Fu Yue (HR 6630) was a lowly wall-builder promoted to high office by the Shang King Wuding. The Weaver Maiden and the Oxherder were said to be a pair of lovers, separated by the Silver River, i.\,e. the Milky Way. For divinatory purposes, a relatively rough identification of the location of any given phenomenon was enough to tie it into a network of meaning. While court astronomers likely did their utmost to create records of any given guest-star that might have suddenly appeared in the heavens, problems including inclement weather, warfare, and limited staffing of astronomical facilities all might have prevented records from being created or preserved to the end of a given dynasty. When the records were compiled into imperial annals or technical treatises, usually in the following, retrospective historical concerns may have shaped with records were included and which were excluded, and ultimately lost.\footnote{This paragraph: personal communication by Jesse Chapman (Department of East Asian Languages and Cultures, University of California-Berkeley).}

 A clustering of guest star records in Fig.~\ref{fig:overview} can be seen in the asterisms forming the Four Holy Beasts and, thus, naming the lunar mansions and at the edge of the asterism of the Purple Forbidden Palace (in the Milky Way region of Cassiopeia). That suggests that not all clusters and voids of the records mapped in Fig.~\ref{fig:overview} allow conclusions concerning the distribution of astrophysical objects (like cataclysmic variables or variable stars at all). Some of them originate from the filters of preservation and divinatory relevance.

\subsection{Our strategy}
 In this paper we present a new method to derive the transient's position using some appropriate cases from the already published collections of nova records by the authors mentioned above. The result of a careful re-reading of the respective historical records and an independent interpretation of the positions given therein (with regard to the existing works on the identification of old asterism names) is a list of positions which is less misleading and ready to use. In this paper, we present only a shortlist drawn out of these existing lists. The full catalogue will be published as database but we use this shortlist to explain the method.
 \begin{figure}
	\includegraphics[width=\columnwidth]{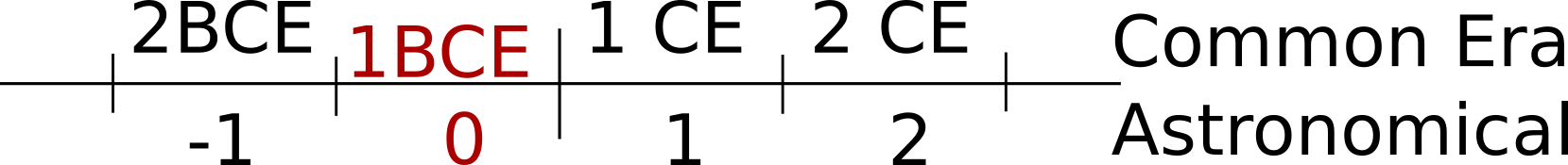}
    \caption{Astronomical Year Counting against year counting of the Common Era.}
    \label{fig:times}
\end{figure}

 \textbf{Wording.} Concerning the descriptions of positions in historical reports we want to point out that there are no coordinates before the 17th century but only verbal descriptions with regard to Chinese asterisms (cf. \citet{hoffmann2019}). A Chinese asterisms can either be one single star or a group of stars which we would name a constellation in western cultures). Hence, in this article (like in several contexts of historical astrometry and sky cultures), we use the term `asterism' as a more general term addressing both the constellations and the single star asterims.

\textbf{Astronomical Year Counting.} Please remember, that Astronomical Year Counting (AYC, with negative numbers) deviates from the counting of years in the Common Era (CE), because AYC counts a year `0' while CE does not (Fig.\ref{fig:times}).
 
 \begin{figure*}
	\includegraphics[width=.97\textwidth]{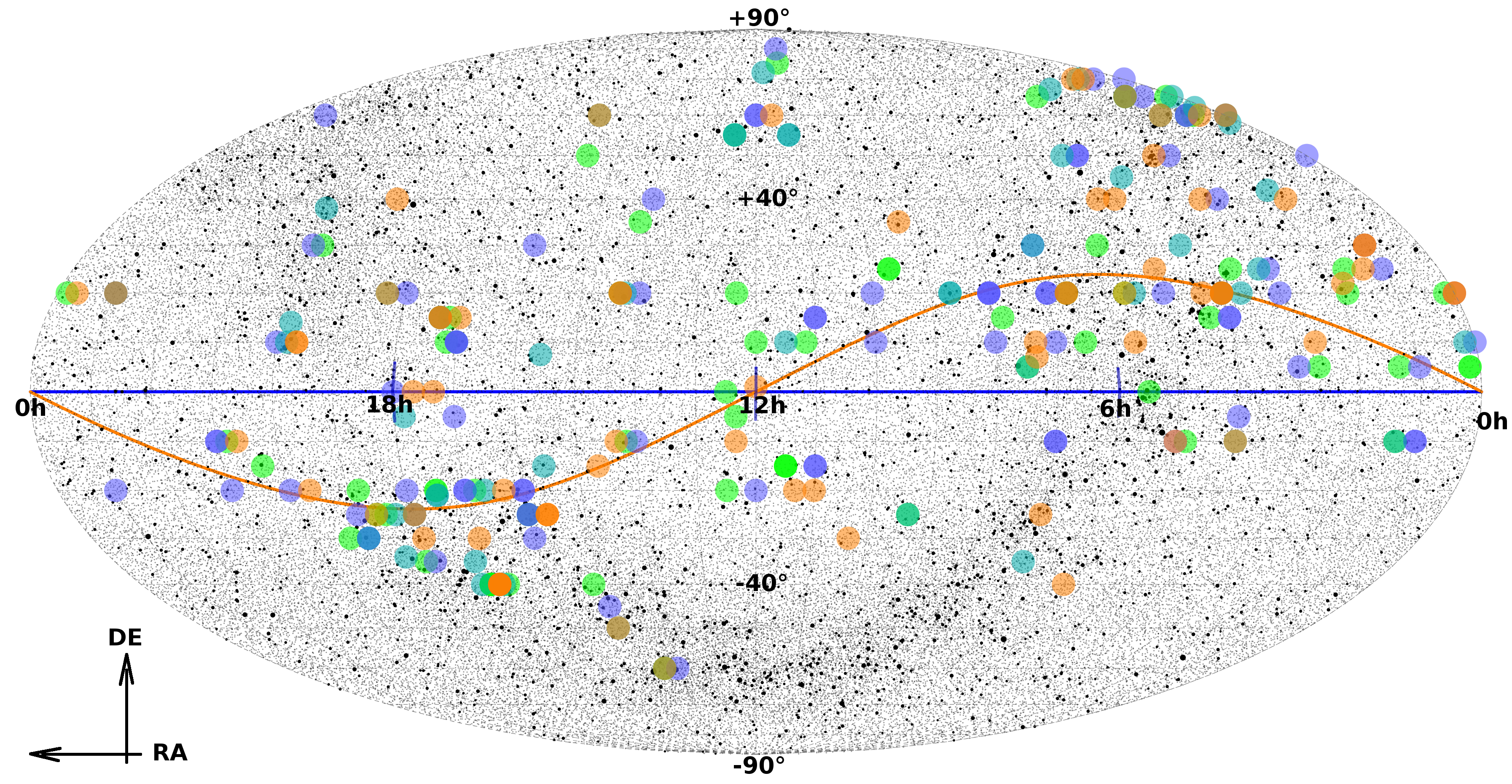} \\
	\includegraphics[width=.97\textwidth]{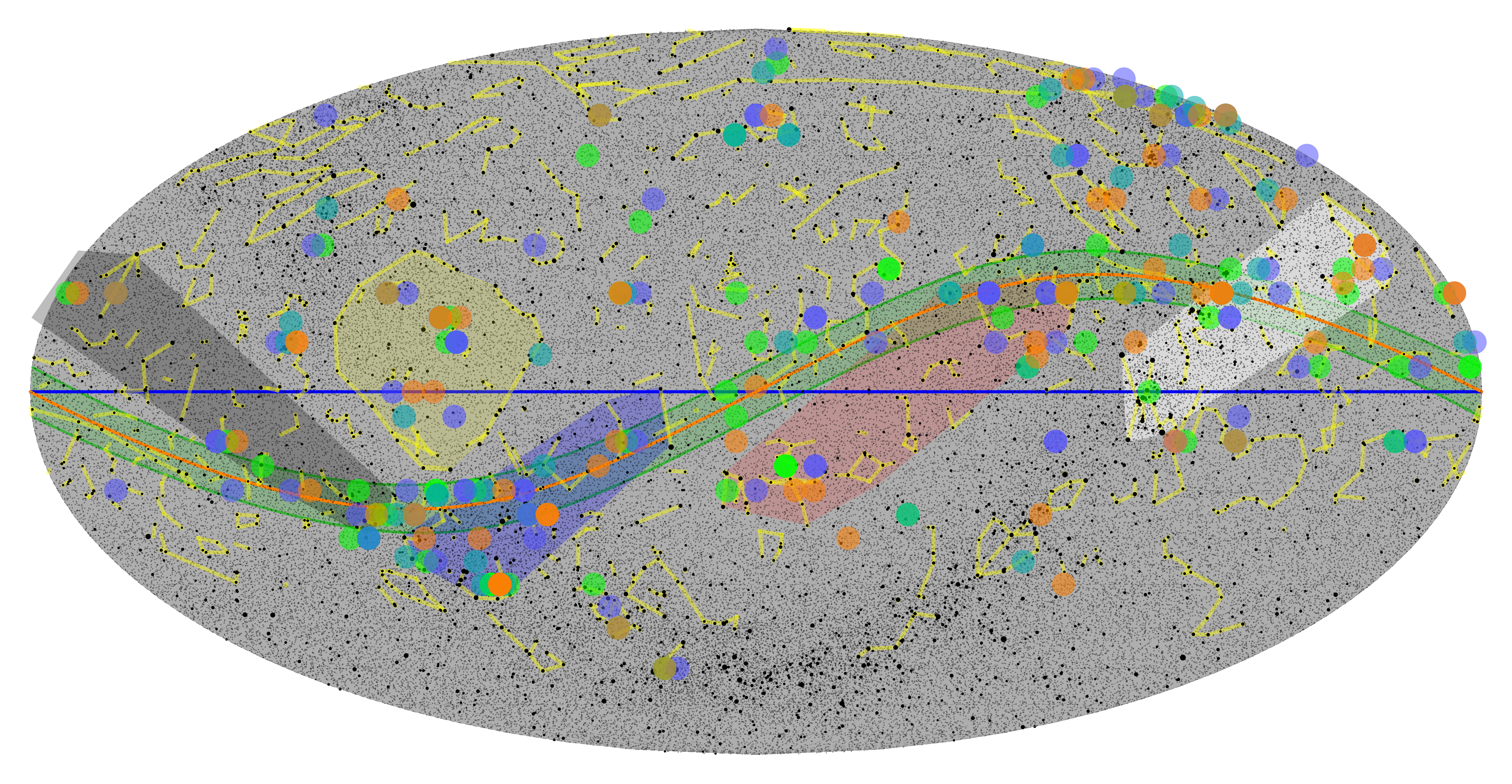}
    \caption{Coordinates of historical novae according to Hsi, Xi and Po, Pskovskii and Stephenson$+$ (indicated by different colours as in Fig.~\ref{fig:litVglN} and \ref{fig:litvgl}), covering $-6$th century to $+17$th century. Equatorial coordinates, ecliptic as sine (green stripe: range of the path of the moon), background star catalogue: HIPPARCOS up to 10~mag and Yale Bright Star Catalogue. The upper map displays the higher star density in the Milky Way, the below map displays the Chinese asterisms (lines) and highlights the Celestial Market Place (yellow) Four Holy Beasts (Dark Warrior, Azure Dragon, Vermillon Bird, White Tiger shown in dark grey, blue, red, light grey, respectively), indicating astrological purposes. The coloured circles represent the (exaggerated) point coordinates of the recorded transients as given by the above authors. The density of records is higher in the galactic bulge and in the Cas-region of the Milky Way (above map) but not in the regions of Orion or Cygnus (both also in the Milky Way). A higher abundance of records even outside the Milky Way, i.\,e. in areas of few stars, occurs along the ecliptic (centre of map) or rather within the Four Holy Beasts and in the Celestial Marketplace. In the southern Milky Way, there are gaps of observation because these parts are invisible from China. See text for further discussion.}
    \label{fig:overview}
\end{figure*}

\section{Localization of historical sightings}
 \subsection{Step 1: Extracting a shortlist of most likely novae and best positioned guest stars.} From three published corpora of historical reports suggested as novae (\cite{hsi}, \cite{ho}, \cite{xu2000}) we selected the entries of most information given in three steps. 

\textit{Selection 1:} While Xu's list already preselected point-like appearances considered to be novae, Ho's list also includes comets. Thus, we first have to exclude all reports mentioning a movement or a tail. We do not consider the terminology as `fuzzy star' (bo xing) alters `guest star' (ko xing); both terms are accepted to designate novae as point-like objects. However, we look more carefully in case the text reports a `broom star' or `sweeping star' (hui xing) because sweeping can imply an extension or a movement. Yet the wording of the description can suggest a movement although there is none; e.\,g.: The event reported in 1006 was a supernova and, thus, always stood at the same place among the stars. However, the text from Song shi\footnote{ch. 56 cited by \cite[p.137, (3)]{xu2000}.} describes this guest star as `passed through the east of Kulou' and another record from Japan (Ichidai y\=ok\=i)\footnote{no. (6) in \cite[Ibid.]{xu2000}.} uses the term `entered Qiguan'. Both wordings could imply a movement although we know that the supernova did not move. Hence, the criterion of selecting records without movement turns out to be physically logical but grammatically not appropriate: The chance to skip fixed guest stars because the terminology suggests a movement is high. Thus, our resulting selection will certainly not be final and complete but it will be a representative shortlist. 

 \textit{Selection 2:} From this (first) shortlist of roughly 180 events only containing appearances with no movement and no tail, we select $i)$ the entries with a precise position (close to a single star and not only a constellation), and $ii)$ the entries with a given duration. As a result we obtain a list of 34 historical events. 

 In a further step of text analysis (\textit{Selection 3:}), we consider the terminology and the description in more detail and we consider the position of the described asterism with regard to the local horizon (visible during the whole night or only shortly in twilight, morning or evening, circumpolar\dots) including the position and phase of the moon. This leads to a grouping in two categories: category~1 are events we consider to be a point source, while in category~2 the description of the transient (e.\,g. a `hui' in twilight pointing somewhere) suggests that it was not point-like but had an extension and was very likely a comet. Even this criterion is as crude as the criteria of the missing movement mentioned above: In some cases, the supernovae are described with long rays and horns\footnote{e.\,g. SN\,1054, record from China (Song huiyao jigao), ch. 52.} implying a fuzzy or extended appearance or even making this explicit\footnote{e.\,g. SN\,1054, record from Japan (Ichidai y\=oki), tei.}. This extended appearance is, of course, caused by turbidity in the atmosphere. However, novae are usually fainter than supernovae which reduces this atmospheric effect. Therefore, we hope that we have largely avoiding incorrectly excluding events based on their fuzzy appearances..  

 Summarizing, we have filtered the whole corpus of almost 600 records selected from the few surviving records in Far Eastern chronicles and published by the three authors mentioned above in the following steps: 
 \begin{enumerate} 
  \item appearances with no explicit tail and no explicit movement (list of 183 entries remaining) 
  \item those of them with most information (34 remaining): 
	\subitem -- a given duration which is longer than one night 
	\subitem -- given with a precise position (referring to a single star) 
  \item those out of this selection where the textual description does not suggest a comet (e.\,g. the verbum being `pointing': 27 remaining). 
 \end{enumerate} 
 Among them, we also find the appearances of the years 369, 393, and 396. A problem concerning the 369 sighting is the huge range of the position: It is given somewhere at the western wall of Zigong palace spanning a half ring around the pole, 19\degr\ to 32\degr\ in declination and right ascensions from 0:49 to 15:25 h. This position is exceptionally uncertain and, therefore, atypical. Additionally, the events in 369 (described above) and in 393 have been suggested as supernovae\footnote{\cite[p.\,176--184]{stephGreen}.} which is also a good reason to consider them as uncertain and avoid including them for the first version of our nova list.

 \textbf{Special case `Nova' 396.} In 396 there was a `large yellow star' which appeared in summer between the Hyades and the Pleiades. It is reported to have disappeared and then reappeared in winter. \cite{hertzog1986} suggests this as an outburst of the white dwarf-red dwarf binary system V471 Tau. Yet, this report appears suspicious in several aspects:
  \begin{itemize} 
   \item The Pleiades are easily recognizable and have been used for many cultural purposes.
   \item The position is remarkably significant. The so called `Golden Gate of the Ecliptic' between two bright star clusters in a way `frames' any appearance in between even for untrained observers. 
   \item The record reports (i) a comet and (ii) a bright object (large yellow star) shortly before its appearance at this remarkable `gate'. They do not report a `guest star' but a bright object which disappears and reappears. 
   \item The dates of the appearance and re-appearance are not exactly given but in the time interval before the (given) date of the appearance of the comet, planet Venus was traveling from the Pleiades eastward in the gap between the Pleiades and the Hyades. The large yellow star must have been next to Venus (at least for a while or at one certain day) and we could wonder why they did not mention Venus to allow a much more accurate positioning.
   \item Interestingly, the time of the invisibility of the 'large yellow star' fit the dates when this planet stood next to the sun and was invisible. Could it be that the `large yellow star' actually \textit{was} planet Venus and not a guest star next to it? 
   \item Of course, Venus is usually white in colour and not yellow but on the one hand, even bright planets like Venus can appear in all shades of yellow, orange or even reddish through a humid atmosphere with smog\footnote{One of us (SH) saw Jupiter reddish while standing almost in zenith when observing from within an Indonesian metropolis.} and on the other hand, human's usage of colour words for the same thing might vary a lot. 
   \item There is zero possibility that the Chinese would mistake Venus for another body in this period: The Chinese assiduously tracked Venus from at least 500 years earlier. The Mawangdui `Five Planets' manuscript (entombed 168 BCE) gives the synodic period for Venus within half a day of the modern value. Mathematical astronomy continued to advance in the ensuing centuries, and it is difficult to imagine that astronomers would have made such an error.\footnote{Thanks to Jesse Chapman in personal communication.} 
	\item However, a chronicler some decades or generations later might have wrote up his report of the dynasty intentionally suggestive with regard to celestial omens and the politics he denoted. Note, that the chronicle does not speak of a guest star but simply notes that `a large yellow star emerged in the space of Mao' and `the yellow star appeared again' \citep[p.\,133]{xu2000}.  
  \end{itemize} 
  
 Due to its uncertainty of being an observation of a classical nova, we decided to not consider the event in our shortlist of primary nova candidates. 
 
 \textbf{The obtained shortlist.} Neglecting the events that are as uncertain as such, we obtain a list of remaining 24 potential novae as a shortlist (texts and localization cf. Tab.~\ref{tab:textselection}) to develop an algorithm for the suggestion of post-nova objects. 

 Among them, the reports from our year 1592 are special because of its complex description. We interpret this as two guest stars without movement in two different constellations observed almost simultaneously. Thus, the list turns out to contain 25 potential classical novae in 24 years. 

 Note, that in all these cases no brightness or even change of brightness is given. Hence, it is impossible to estimate a light curve. 
 
 \begin{table*}
	\centering
	\caption{A representative subset of historical records which might witness classical nova outbursts. It is highly likely that the `broom stars' are still comets and therefore not in Xu's collection. (Ho in the 1960s used another transliteration.)}
	\label{tab:textselection}
	\fontsize{9}{10.8}\selectfont
	\begin{tabular}{rp{.22\textwidth}lp{.2\textwidth}p{.22\textwidth} }  
		\hline
	year	&extracted information from text	&cited from &position given by	&where \\
		\hline
-203	&fuzzy star in Dajiao	&Xu et al. (2000)	&single star-asterism	&$\alpha$ Boo\\
-103	&fuzzy star at Zhaoyao	&Xu et al. (2000)	&single star-asterism	&$\gamma$ Boo\\
-47	&about four chi east of the second star of Nandou	&Xu et al. (2000)	&star in constellation	&east of $\lambda$ Sgr\\
-4	&broom star at Chhien-Niu	&Ho (1962)	&constellation	&at $\rho, o, \pi, \beta, \alpha, \xi$ Cap\\
64	&guest star outside the Nanduan gate of Taiwei, near Zuozhifa	&Xu et al. (2000)	&single star asterism	&near Zuozhifa $=\eta$ Vir\\
70	&guest star in Xuanyuan	&Xu et al. (2000)	&constellation	&in chain of $\rho, o, 31, \alpha, \eta, \gamma, \zeta, \mu, \epsilon, \lambda, \kappa, f$ Leo, HIP47617 $\alpha, 38$ Lyn, HR 3612, 10 UMa\\
101	&guest star in space of the fourth star of Xuanyuan	&Xu et al. (2000)	&star in constellation	&$\alpha$ Lyn or $\zeta$ Leo \\
329	&fuzzy star in the northwest trespassing against Dou	&Xu et al. (2000)	&constellation (northern Dou) &(Bei)Dou: $\alpha, \beta, \gamma, \delta, \epsilon, \zeta, \eta$ UMa\\
641	&fuzzy star in Taiwei, trespassing against Langwei	&Xu et al. (2000)	&2 constellations	&Taiwei is an enclosure, Langwei north of it (Coma Berenices)\\
667	&broom star among Wu-Chhê and Pi and Mao	&Ho (1962)	&3 asterisms	&Mao = Pleiades, Pi = Hyades, Wu-Chhê = $\iota, \alpha, \beta, \theta, \gamma$ Aur\\
668	&broom star above Wu-Chhê	&Ho (1962)	&constellation	&above $\iota, \alpha, \beta, \theta, \gamma$ Aur\\
683	&broom star north of Wu-Chhê	&Ho (1962)	&constellation	&north of $\iota, \alpha, \beta, \theta, \gamma$ Aur\\
722	&guest star beside Gedao	&Xu et al. (2000)	&constellation	&beside $o, v, \theta, \phi, \delta, \epsilon, \iota$ Cas\\
840	&broom star between Ying-Shih and Tung-Pi	&Ho (1962)	&2 asterisms	&between $\gamma$ Peg--$\alpha$ And and $\alpha, \upsilon, \beta$\dots Peg\\
891	&guest star east of Dongxian star, distance about one cun (0.1 deg)	&Xu et al. (2000)	&Dongxian is constellation, but Dongxian star likely designates its principle star	&$\psi$ Oph\\
1175	&fuzzy star in northwest, outside the wall of Ziwei + above Qigong	&Xu et al. (2000)	&2 asterisms	&Qigong $=\beta$ Boo + 6 stars, Zigong $=$ circumpolar area, starting at $\theta, \iota, \alpha$ Dra + 12 stars\\
1430	&guest star more than one chi northeast of Nanhe	&Xu et al. (2000)	&constellation	&$alpha, \beta, \epsilon$ CMi\\
1431	&Hanyu star like a crossbow pellet beside Jiuyou	&Xu et al. (2000)	&constellation	&$\mu, \omega, 63, 64, 60, 58, 54$ Eri, 1 Lep, HIP 21515\\
1437	&guest star between 2nd and 3rd star of Wei, nearer to 3rd, separated by about half a chi	&Xu et al. (2000)	&stars in constellation	&3rd star of Wei $=\zeta$ Sco, 2nd star of Wei: $\mu$ or $\epsilon$ Sco\\
1461	&a star as white as powder by the side of Tsung-Chêng in the Thien-Shih	&Ho (1962)	&constellation	&$\beta, \gamma$ Oph\\
1497	&guest star beside the Tianji star	&Xu et al. (2000)	&Tianji is constellation, but Tianji star likely designates its principle star	&$\beta$ UMi\\
1592	&guest star east of Tiancang (3 cun east of 3rd star?); a guest star east of Wangliang	&Xu et al. (2000)	&2 asterisms: 1 constellation + 1 single star(?)	&Tiancang $=\iota, \eta, \theta, \zeta, \tau, 57$ Cet, 3rd star $=\theta$ Cet, Wangliang $=\beta, \kappa, \eta, \alpha, \lambda$ Cas\\
1661	&guest star in Nü	&Xu et al. (2000)	&constellation &	$\rho, o, \pi, \beta, \alpha, \xi$ Cap\\
1690	&anomalous star east of the 3rd star of Ji	&Xu et al. (2000)	&single star-asterism	&$\epsilon$ Sgr\\
		\hline
	\end{tabular}
\end{table*} 
 
 \subsection{Step 2: Defining fields to search for post-novae.} 
 The obtained list of historical sightings consists of appearances with no movement and no tail reported, and the duration explicitly given longer than one night or the position given next to a single star asterism. For these events we were analyzing the position with regard to the fixed stars.

 All texts follow the scheme 
 \begin{verse} 
 name of Emperor, $n$-th year of reign, [$m$-th month,] [$d$-th day] \\ `A ( ) star appeared at' name of asterism 
 \end{verse} 
 The day and the month are not always given and, therefore, written in brackets. In some cases, the text also mentions a date of disappearance or a duration (like `lasting 185 days' for the supernova in 1181). The brackets [] indicate the possible specification of the star as `new star', `guest star', `broom star', `fuzzy star' and such -- in case of the known supernovae they are usually termed `guest stars' but all other appearances (classical nova outbursts, tailless comets, meteors, maybe variable stars of Mira type and other M-type giant stars with possible long-term variations in their pulsation properties, maybe even more supernovae etc.) are designated with a variety of terms. 

\begin{figure}
	\includegraphics[width=\columnwidth]{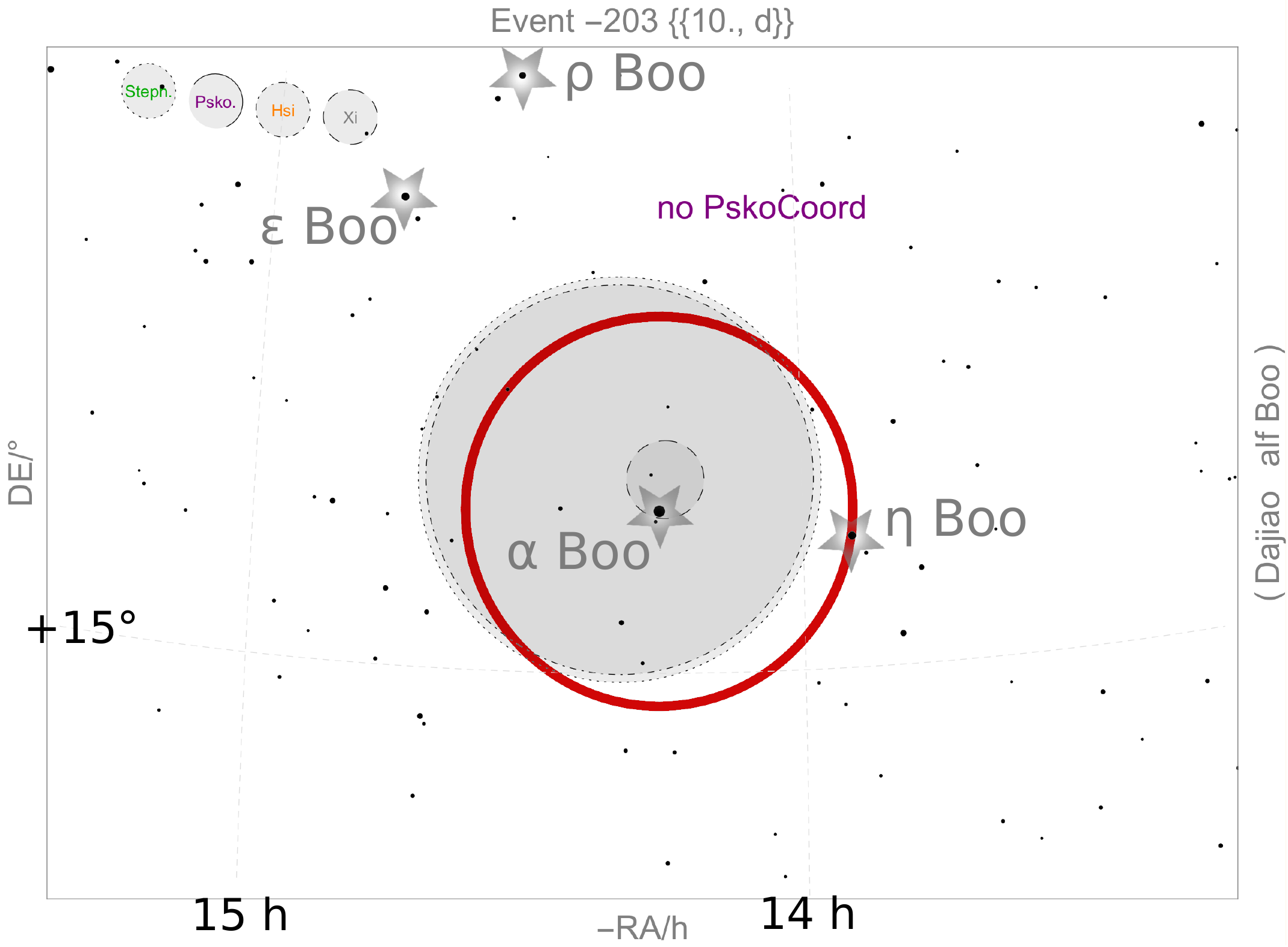}
    \caption{Example for a given single star asterism: Coordinates of historical novae $-203$ according to the lists of \citet{hsi}, \citet{xi+po}, \citet{pskovskii}, and \citet{stephenson}: Four authors evaluating the same historical texts to derive positions of the phenomenon obtain almost the same results next to the bright star Arcturus.}
    \label{fig:event203}
\end{figure}

\begin{figure}
	\includegraphics[width=\columnwidth]{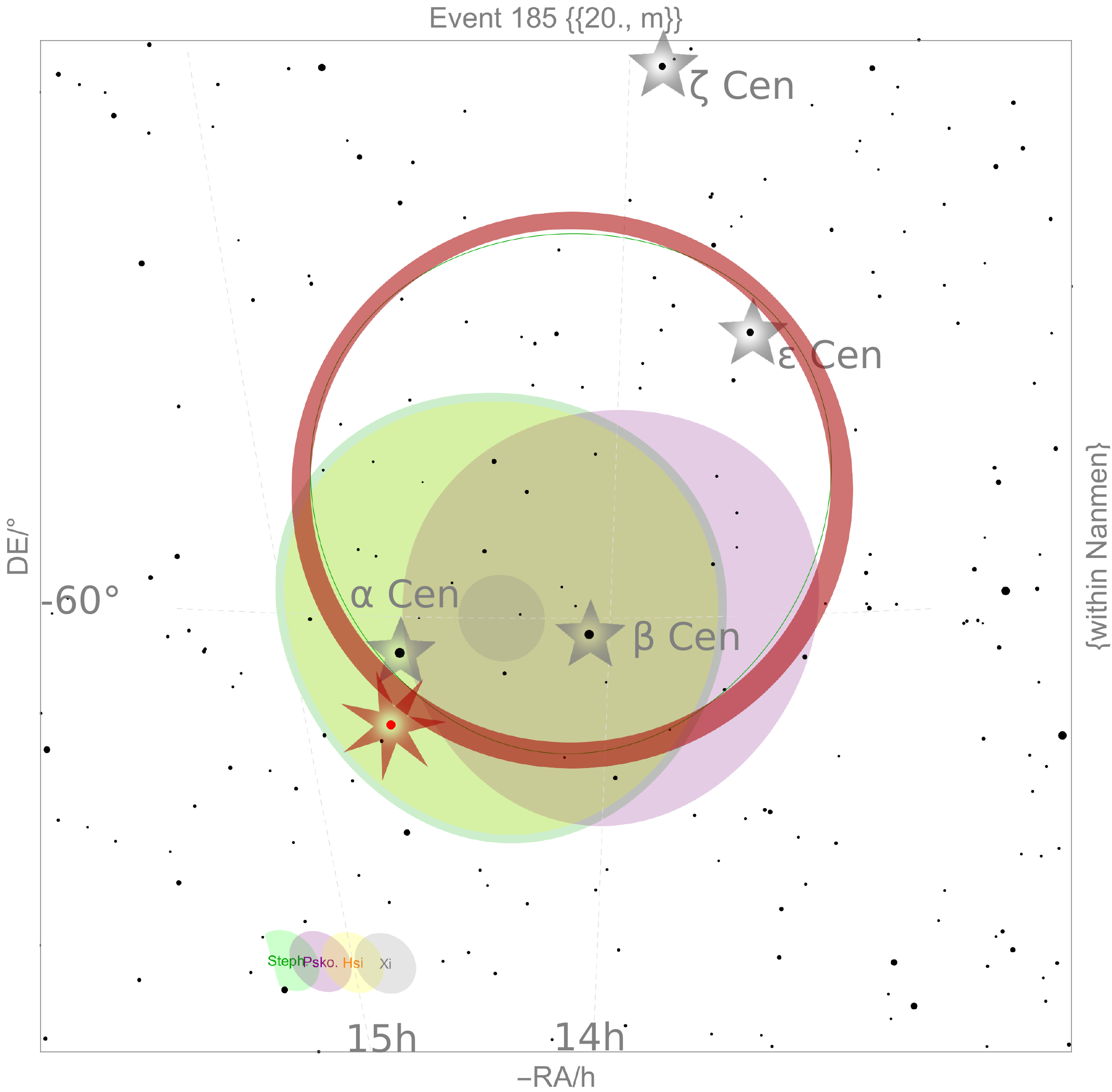}
    \caption{Another (more complicated) example for a given single star asterism: The positions the considered authors define for the historical event: Stephenson, Pskovskii, Hsi, Xi and Po; green, purple, yellow, gray respectively. The position of the associated SNR is marked with a red dot and highlighted by a star with seven spikes.}
    \label{fig:event185}
\end{figure}

\begin{figure}
	\includegraphics[width=\columnwidth]{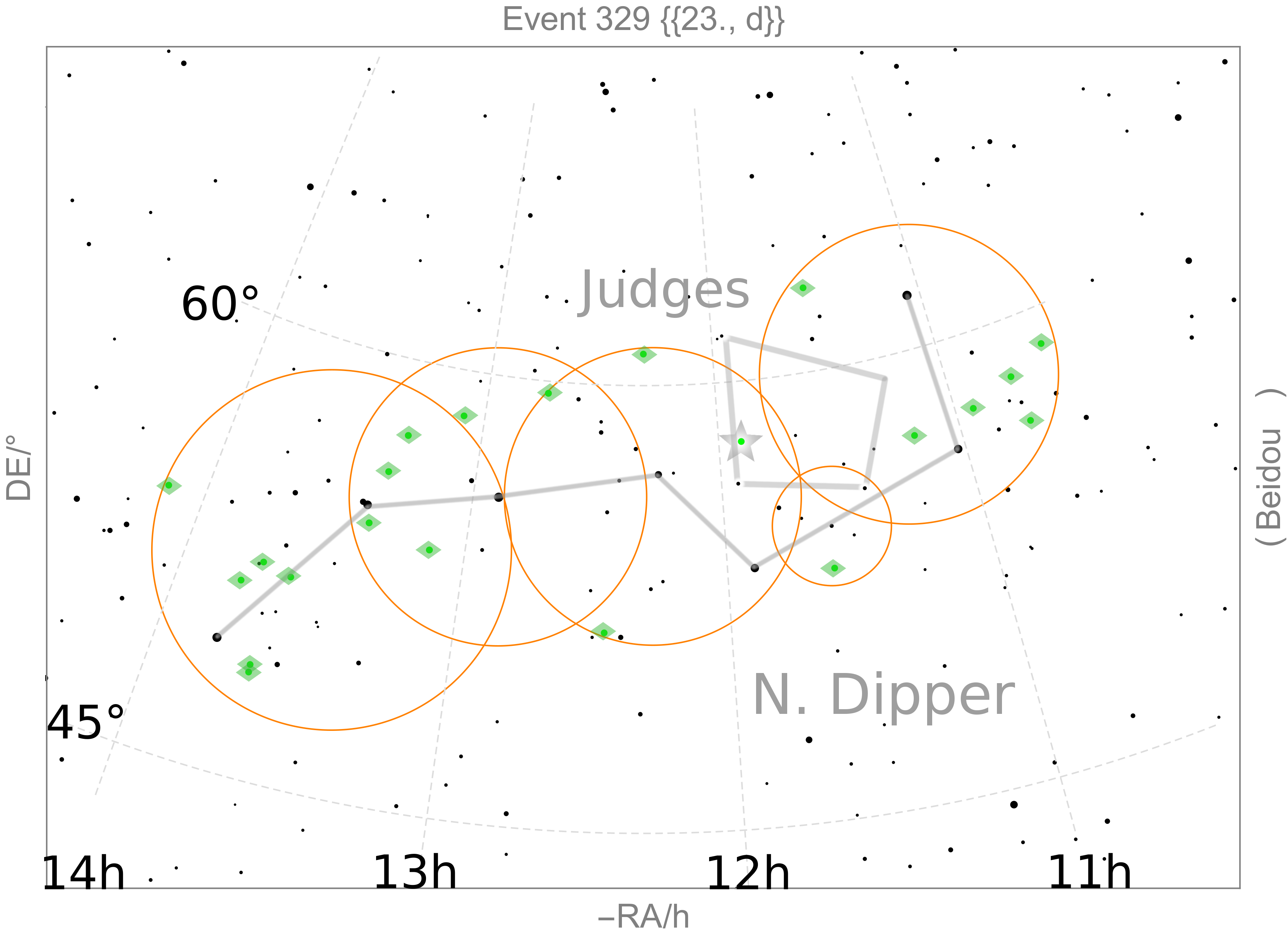}
    \caption{Example for a given constellation and the result of the graphical analysis: orange circles for the fields to search candidates, green dots are known CVs (2018). The one CV within our circles of search which is highlighted with a star symbol has to be thrown out because it is within the asterism of the Judges for Nobility. The record reports that the transient object `trespassed against [Bei-]Dou', implying it was likely closer than 1\degr\ to the line of the Dipper.}
    \label{fig:event329}
\end{figure}

\begin{figure}
	\includegraphics[width=\columnwidth]{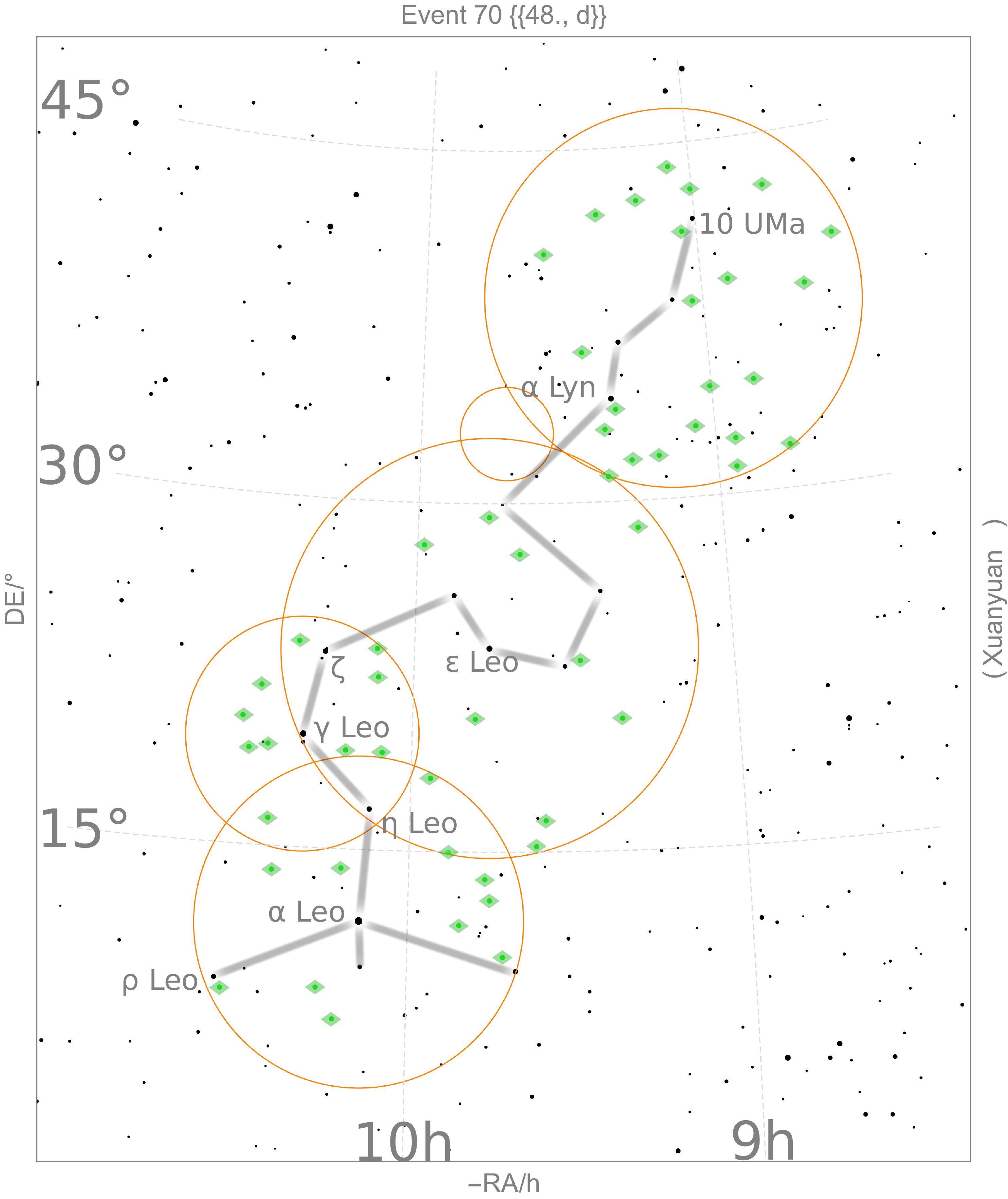}
    \caption{A less complicated example for a given constellation: A more typical graphical analysis is the transient in year 70 CE in the asterism Xuanyuan: The grey lines are the lines of the asterism as drawn on the preserved historical Suzhou map (cf. \citet{rufus}, \url{http://www.chinesehsc.org/zoomify/suzhou_planisphere.html}). The orange circles which cover the constellation have different sizes and are centered in appropriate HIP stars. The green dots indicate the known CVs (2018) which fit the given position.}
    \label{fig:event70}
\end{figure}

 The typical steps which, then, have to be performed are:  
  \begin{enumerate}
	\item Find out what is designated by the asterism name. The translation of the term does not matter, it is only important which (group of) star(s) is designated with it. 
	\item Find it on a map and find out the positions of neighbouring asterisms. 
	\item Where was the appearance seen?
	\subitem $\bullet$ \textit{within a constellation}: define circles to cover the area of the constellation (e.\,g.: Fig.~\ref{fig:event70} for a typical case, Fig.~\ref{fig:event329} for an exceptional case with two asterisms nested in each other). 
	\subitem $\bullet$ \textit{next to a single star}: define a small circle around the given position (if it is given as distance to a star like `$n$ \textit{chi} east of a star') or around the star. The radius of the circle is chosen either to cover the given distance or (in case of no given distance) as roughly half the distance to the neighbouring asterisms (e.\,g.: Fig.~\ref{fig:event203} for a typical case, Fig.~\ref{fig:event185} for an exceptional case with a hardly identifiable star). 
	\item Check the position of the asterism with regard to the horizon at day (or month) of appearance. In most cases, this does not matter but in some cases, this might lead to further (alternative) fields of search: e.\,g. in year 668, there is a report `broom star above Wu-Chhê', the description `above' instead of terms like `north', `east', `west' and alike suggests an interpretation with regard to the local horizon. The asterism `Wu-Chhê' ($=$Wuche) consists of the bright stars of Auriga which rises and sets in China. Thus, the area `above' is different in the morning and evening which is why we give two alternative fields of search for this event in Tab.~\ref{tab:tinynovacat}. 
 \end{enumerate} 
 
 The following two examples shall demonstrate our method (with no dependency on the horizon). First, we will explain the method for the very clear case of the appearance in $-203$, then we apply this method on descriptions of the supernova n $185$. This method of defining the circles of search according to the historical record is the same but in the first case, the asterism is clearly identified, in the second case it is not.   
 
 \textbf{Example of Appearance in $-203$}: The preserved record is translated by \cite[p.\,129]{xu2000} ``3rd year of emperor Gaozu of Han, 7th month. A fuzzy star (xing bo) appeared in Dajiao for over ten days before going out of sight.'' (Han shu, Wuxing zhi, ch. 27). The variations in the translations by Ho and Hsi concern only the verbum (to appear, to be seen) not the position in the sky, which is why we do not arbitrate between them. After ensuring the reliablity of the translation we performed the four steps listed above:  

 Dajiao (the Great Horn) is the name of the single star asterism of $\alpha$~Boo (Arcturus)\footnote{According to \cite[p.\,422]{xu2000}, \cite[p.\,17; item 31. in Blue Dragon]{rufus} and others}. The neighbouring asterisms are the Left and the Right Conductor, the Boats and Lake, and the Mattress of the Emperor with distances of 3\degr\ to 6\degr. Hence, we choose to search for modern counterparts of this event within a circle of 3\degr\ around Arcturus (cf. Fig.~\ref{fig:event203}). 

 \textbf{Example of SN 185}: In some cases, it is not as clear as in the above example how to identify the designated star(s). That is why we also describe the case of the supernova in 185. The historical record reads like this: ``Emperor Ling of Han, 2nd year of the Zhongping reign period, 10th month, day guihai [60]. A guest star emerged within Nanmen. It was as large as half a mat, with scintillating variegated colors. It grew smaller and in the 6th month of the year after the next it disappeared.'' \cite[p.\,131]{xu2000}. 

 In this case, the asterism Nanmen (the Southern Gate) is not a single star but a constellation of two stars in Centaurus. According to \cite[p.\,17]{rufus} it is typically identified with $\alpha$ and $\beta$ Centauri while according to more recent literature \citep{xu2000} the main star is $\zeta$ Cen. Summarizing these and other suggestions, the Southern Gate can be identified to be the pairs of ($\alpha$ and $\beta$), ($\alpha$ and $\epsilon$), ($\epsilon$ and $\zeta$), ($\alpha$ and $\zeta$)\dots\ and there may be further options. In this case, we consider the star HIP\,68992 as the centre of the circle with a radius of 6\degr\ for including at least $\alpha$, $\beta$, and $\epsilon$ Cen (see Fig.~\ref{fig:event185}). If we included also $\zeta$ Cen, the circle should have a diametre of 16\degr\ (which is the distance between $\alpha$ and $\zeta$) but this combination appears unlikely due to the huge distance between the stars (cf. Fig.~\ref{fig:event185}). 
 
 \textbf{Applying the procedure to our shortlist.} In our shortlist of nova candidates, the asterisms are more clearly identifiable. However, in 17 of 28 cases, the position is not given by a single star but by a constellation and in some cases of given single stars their identification is uncertain. Looking for candidates of nova- and supernova-remnants (or variable stars or any other type of point-like phenomenon) in cases of positions given in a constellation we have to search within the whole constellation. The current database for variable stars (like cataclysmic variables) is the International Variable Star Index (VSX, \citet{watson}) provided by the American Association of Variable Star Observers (AAVSO). As this index offers to request circles (or boxes) surrounding a coordinate, we covered the area of the constellations with circles. 

 For instance, in case of the event in the year 329 the text reports `There was a fuzzy star [bo xing] in the NW trespassing against Dou. After 23 days it was extinguished.' \cite[p.\,133]{xu2000}. There are two constellations named Dou but only one of them can be seen in the northwest. This is the asterism similar to our Big Dipper (the other Dou would be in Sagittarius). The term `trespassing against' has the astrometrical meaning to approach the asterism to within one du where 1~du$=360/365.25=0\fdg9856\approx1\degr$. That means, we first define circles covering the whole constellation, search for CVs and symbiotic stars (as nova candidates, while in case of supernovae we would search for pulsars and SNRs in this field) in the current catalogue. Afterwards, we plot the findings and select those of the objects which fit best the given description (in this case: close to the grey line connecting the stars of the Big Dipper).   

 The size of the circles was chosen to cover the asterism given in the description. In case the report preserves the information on a guest star only somewhere within a constellation (not within one du or other units), the circles should cover the area of this constellation plus at least half of the distance to the next constellation (because otherwise, the guest star would have likely been reported in the other asterism). In case of single stars like `near Zuozhifa' ($\eta$~Vir) for the appearance in year 64~CE the radius of our circle is normally assumed as 3\degr\ but the objects found within this radius have to be reconsidered with regard to the historical report: In case the single star is a single star asterism (like The Great Horn equaling Arcturus) we apply the normal rule to choose a radius as half distance to the neighbouring asterisms and achieve e.\,g. 5\degr\ for the circle of guest star in $-203$ (`in Dajao' while Dajao designates the single star Arcturus). In case the single star is part of a constellation (like `3rd star of Wei' as reported in 1437), we maybe have to vary the rule: e.\,g. take half the distance to neighbouring stars (of the same constellation) or argue for other criteria in special cases. 

  \subsection{Step 3: Identifying possible candidates (Outlook)}
  The next step should be the search for modern counterparts in the fields identified above which is, however, beyond the scope of this paper. The difficulty to securely identify a certain object as the cause of the historical appearance will be the uncertainty of the nature of the appearance and the huge arrays in which we have to search. First, we will be looking for CVs in the fields defined above in order to identify candidates of possible classical novae. However, in none of the cases we can derive a light curve from the historical reports and in most cases we do not even know the apparent brightness, duration, and colour. That is why, it always remains an uncertainty that this exclusive search for CVs is allowed. Second, there is no additional criterion to exclude certain CVs in the field: As records in many cases only report the constellation it is generally not allowed to select a certain star (e.\,g. main star) or the vicinity of a constellation line to rule out some candidates within the defined fields. Astrophysical criteria such as the distance of the CV also do not apply because the important question is only whether or not the current brightness of the object and the (known) amplitude(s) of classical novae allow to brighten the CV to naked eye visibility.
 
 \textbf{CVs in Figs.~\ref{fig:event329} and \ref{fig:event70}:} Of course, in huge areas like the area of a whole constellation, there are many CVs. In Figs.~\ref{fig:event329} and \ref{fig:event70} the green symbols show all CVs listed in the VSX catalogue with coordinates inside our circles of search.\footnote{This query has been performed in September 2018 and, in the meantime, there are some additional objects listed. However, the new contributions, especially the faint sources discovered by GAIA Survey are not appropriate to fulfill the conditions of causing historical novae.} Most of them are not valid counterparts for ancient classical nova events. One has to consider carefully the physical qualities of each CV and discuss whether or not the certain system would be able to produce an eruption visible to the naked eye. Typical amplitudes of classical novae are 11 to 13 mag but the maximum amplitude observed during the last 1.5 centuries is 16 mag (cf. Fig.~1 in \citet{vogt2019}). Assuming optimistically a (faint) detection limit of 5~mag appearances for naked eye observers (which certainly does not apply at least in regions of bright celestial background as in the clouds of the Milky Way: cf. \citet{hoffmann2019}, Fig.~12 for the error or estimating stellar magnitudes, and Fig.~\ref{fig:distrMag}.) a CV candidate should have a current apparent brightness of $5+16=21$~mag. Beyond this limit, we consider the system as impossible and between 18 and 21~mag as possible but unlikely candidate to cause a naked eye nova. In a next step, we discuss the remaining candidates, which is yet beyond the scope of this paper. We are going to present a list of valid candidates soon in another paper (Hoffmann and Vogt, in preparation).   
 
 \begin{table*}
	\centering
	\caption{List of possible historical novae and their positions: The positions and radii are ready to use in catalogue queries but it is possible the output will be correct but not the final result (as shown in Fig.~\ref{fig:event329}). It should be checked again with a map of Chinese asterisms and of course checked regarding the physical qualities of the obtain catalogue members whether or not the object might have caused the historical transient. The coordinates given to equinox 2000, we recommend to rely only on the HIP star number.}
	\label{tab:tinynovacat}
	\fontsize{8}{9.6}\selectfont
	\begin{tabular}{rrrlp{8em}lrrrrp{15em}} 
		\hline
		yr & ID & dur & dur & asterism & star & HIP & RA & DE & radius & commentary \\
		\hline
 -203. & &  $>$10. & \text{d} & \text{Dajiao} & \text{$\alpha$ Boo} & 69673. & 213.915 & 19.1824 & 5 & \text{} \\
 -103. & \text{} & ? & ? & \text{Zhaoyao} & \text{$\gamma$ Boo} & 71075. & 218.019 & 38.3083 & 3 & \text{} \\
 -47. & \text{} & ? & ? & 4\degr east of 2nd star of Nandou & \text{$\lambda$ Sgr} & 91974. & 281.207 & -25.0109 & 4 & 
   $\lambda$ Sgr is HIP 90496, a star 3\degr 51$^\prime$ east of this is HIP 91974  \\
 -4. & 1. & \text{} & \text{} & \text{Chhien-Niu} & \text{} & 100345. & 305.253 & -14.7815 & 3 & \text{} \\
 -4. & 2. & \text{} & \text{} & \text{Chhien-Niu} & \text{} & 99572. & 303.108 & -12.6175 & 1.5 & \text{} \\
 -4. & 3. & \text{} & \text{} & \text{Chhien-Niu} & \text{} & 100881. & 306.83 & -18.2117 & 2 & \text{} \\
 64. & \text{} & 75 & \text{d} & \text{Zuozhifa} & \text{$\eta$ Vir} & 60129. & 184.976 & -0.666833 & 5 & \text{} \\
 70. & 1. & 48 & \text{d} & \text{Xuanyuan} & \text{} & 44700. & 136.632 & 38.4522 & 8 & \text{} \\
 70. & 2. & 48 & \text{d} & \text{Xuanyuan} & \text{} & 47617. & 145.644 & 32.9949 & 2 & \text{} \\
 70. & 3. & 48 & \text{d} & \text{Xuanyuan} & \text{} & 47908. & 146.463 & 23.7743 & 9 & \text{} \\
 70. & 4. & 48 & \text{d} & \text{Xuanyuan} & \text{} & 49669. & 152.093 & 11.9672 & 7 & \text{} \\
 70. & 5. & 48 & \text{d} & \text{Xuanyuan} & \text{} & 50583. & 154.993 & 19.8415 & 5 & \text{} \\
 101. & 1. & ? & ? & \text{4th star of Xuanyuan} & \text{$\alpha$ Lyn} & 45860. & 140.264 & 34.3926 & 3 & \text{} \\
 101. & 2. & ? & ? & \text{4th star of Xuanyuan} & \text{$\zeta$ Leo} & 50335. & 154.173 & 23.4173 & 3 & \text{} \\
 329. & 1. & 23 & \text{d} & \text{Beidou} & \text{} & 65802. & 202.324 & 53.1776 & 6 & \text{} \\
 329. & 2. & 23 & \text{d} & \text{Beidou} & \text{} & 59882. & 184.209 & 56.2815 & 5 & \text{} \\
 329. & 3. & 23 & \text{d} & \text{Beidou} & \text{} & 54495. & 167.24 & 59.2153 & 5 & \text{} \\
 329. & 4. & 23 & \text{d} & \text{Beidou} & \text{} & 62956. & 193.507 & 55.9598 & 5 & \text{} \\
 329. & 5. & 23 & \text{d} & \text{Beidou} & \text{} & 56510. & 173.77 & 54.7854 & 2 & \text{} \\
 369. & \text{} & 5. & \text{m} & \text{Zigong} & \text{} & \text{} & 0. & 90. & 32 & From this circle we subtract the inner 13\degr DE and take right ascension between 0:49 and 15:25 h. \\
 641. & 1. & \text{} & \text{} & \text{Thai-Wei trespassing Lang wei } & \text{} & 59078. & 181.732 & 20.4929 & 1.5 & \text{southern half} \\
 641. & 2. & \text{} & \text{} & \text{Thai-Wei trespassing Lang wei } & \text{} & 60957. & 187.43 & 20.8961 & 4 & \text{southern half} \\
 667. & \text{} & \text{} & \text{} & \text{among Wu-ch{'}e, Pi, Mao} & \text{} & 21010. & 67.5836 & 28.1319 & 7 & \text{} \\
 668. & 1. & \text{} & \text{} & \text{above Wu-Chh{\' e}} & \text{} & 29949. & 94.5703 & 46.3604 & 3 & \text{evening \textbf{(alternatively)}} \\
 668. & 2. & \text{} & \text{} & \text{above Wu-Chh{\' e}} & \text{} & 22545. & 72.7888 & 48.7407 & 5 & \text{morning \textbf{(alternatively)}} \\
 683. & \text{} & \text{} & \text{} & \text{north of Wu-Chh{\' e}} & \text{} & 26569. & 84.7365 & 49.4162 & 5 & \text{} \\
 722. & 1. & 5 & \text{d} & \text{Gedao} & \text{} & 3504. & 11.1812 & 48.2844 & 5 & \text{} \\
 722. & 2. & 5 & \text{d} & \text{Gedao} & \text{} & 5542. & 17.7757 & 55.1499 & 3 & \text{} \\
 722. & 3. & 5 & \text{d} & \text{Gedao} & \text{} & 6686. & 21.4539 & 60.2353 & 4 & \text{} \\
 722. & 4. & 5 & \text{d} & \text{Gedao} & \text{} & 8886. & 28.5989 & 63.6701 & 2 & \text{} \\
 722. & 5. & 5 & \text{d} & \text{Gedao} & \text{} & 11569. & 37.2664 & 67.4025 & 3 & \text{} \\
 840. & \text{} & \text{} & \text{} & betw. Ying-Shih and Tung-Pi & \text{} & 116527. & 354.235 & 21.7231 & 9 & \text{} \\
 891. & \text{} & ? & ? & \text{Dongxian} & \text{$\psi$ Oph} & 80343. & 246.026 & -20.0375 & 3 & consider only stars in the eastern half. One of three records gives a double hour for the appearance; this could allude to short visibility (meteor?).\\
 1175. & \text{} & 5. & \text{d} & outside the wall of Ziwei and above Qigong & \text{} & 73100. & 224.096 & 49.6284 & 8 & \text{} \\
 1430. & 1. & 26 & \text{d} & \text{Nanhe} & \text{} & 37839 & 116.366 & 7.76447 & 3 & \text{} \\
 1430. & 2. & 26 & \text{d} & \text{Nanhe} & \text{} & 36606 & 112.922 & 10.0736 & 1.5 & \text{} \\
 1431. & 1. & 15 & \text{d} & \text{Jiuyou} & \text{} & 21515 & 69.307 & 0.99833 & 5 & \text{} \\
 1431. & 3. & 15 & \text{d} & \text{Jiuyou} & \text{} & 22701 & 73.2236 & -5.4527 & 4 & \text{} \\
 1431. & 4. & 15 & \text{d} & \text{Jiuyou} & \text{} & 23231 & 74.9822 & -12.5374 & 4 & \text{} \\
 1431. & 5. & 15 & \text{d} & \text{Jiuyou} & \text{} & 22479 & 72.5484 & -16.2172 & 4 & \text{} \\
 1431. & 6. & 15 & \text{d} & \text{Jiuyou} & \text{} & 21763 & 70.1105 & -19.6715 & 6 & \text{} \\
 1437. & \text{} & 14. & \text{d} & \text{3rd star Wei} & \text{$\zeta$ Sco} & 82554 & 253.114 & -40.723 & 2 & north of $\zeta$ \text{Sco $=$ HIP\,82729}. \\
 1461. & \text{} & 3. & \text{d} & \text{by the side of Tsung-Ch{\^ e}ng} & \text{} & 86742. & 265.868 & 4.56733 & 4 & \text{} \\
 1497. & \text{} & ? & ? & \text{Tianji} & \text{$\beta$ UMi} & 72607. & 222.676 & 74.1555 & 3 & \text{} \\
 1592. & 1. & 15 & \text{m} & \text{Tiancang} & \text{} & 8645. & 27.8652 & -10.335 & 1 & \textbf{\textbf{1st event of the year}} \\
 1592. & 2. & 4 & \text{m} & \text{Wangliang} & \text{} & 2543. & 8.09079 & 58.3386 & 5 & \textbf{2nd event}: One could leave aside the inner circle (2$\degr$). \\
 1661. & \text{} & 19 & \text{d} & \text{N{\" u}} & \text{} & 102614. & 311.915 & -7.11112 & 3 & \text{} \\
 1690. & \text{} & 2 & \text{d} & \text{3rd star Ji} & \text{$\epsilon$ Sgr} & 910914. & 278.491 & -33.0166 & 3 &Object east of $\epsilon$ Sgr$=$HIP\,90185. Choose a star $3\degr$ away and search in a circle of
   $3\degr$ radius. \\
		\hline
	\end{tabular}
\end{table*}

\section{Estimating the localization errors using supernovae}
 To estimate the error of the method above we consider the eight historical supernovae for which almost certain identifications exist. Their positions have been derived from historical Far Eastern guest star records by different authors obtaining different results (see Tab.~\ref{tab:SNcoord}). Now we apply our method as explained above and determine the deviation of the position given in the text from the post-supernova object (SNR or PSR). 

\begin{table}
	\centering
	\caption{Coordinates of historical SNe derived by earlier authors (all coordinates RA,DEC in degree and for equinox 1950).}
	\label{tab:SNcoord}
	\fontsize{8}{9.6}\selectfont
	\begin{tabular}{rcccc} 
		\hline
		year & Hsi & Pskovskii & Xi$+$Po & Steph.$+$ \\
		\hline
  185 & $(-215,-60)$ & $(-210,-60)$ & $(-215,-60)$ & $(-215,-60)$ \\
  369 & $(0,60)$ & $--$ & $(-48,67)$ & $--$ \\
  386 & $(-280,-25)$ & $(-285,-30)$ & $--$ & $(-277.5,-25)$ \\
  393 & $(-255,-40)$ & $(-255,-40)$ & $(-260,-40)$ & $(-257.5,-40)$ \\
 1006 & $(-225,-50)$ & $(-225,-45)$ & $(-225,-50)$ & $(-227.5,-40)$ \\
 1054 & $(-85,20)$ & $(-75,20)$ & $(-82.5,20)$ & $(-85,20)$ \\
 1181 & $(-25,70)$ & $(-15,65)$ & $(-22.5,65)$ & $(-22.5,65)$ \\
 1572 & $(-10,60)$ & $(0,60)$ & $(-2.5,65)$ & $(-5,65)$ \\
 1604 & $--$ & $(-255,-20)$ & $(-262.5,-21)$ & $(-262.5,-20)$ \\
		\hline
	\end{tabular}
\end{table}

 As these historical supernovae provide identified modern counterparts we can use them to check the accuracies of the Far Eastern records. One of the big problems is that the preserved records are not necessarily the original reports of professional astronomers but originate from chronicles where they had to fulfill a divinatory function. Later historians might have edited the original records and certainly did not include all the material they found. The records cannot make up a complete set and the context of astronomical reports is almost lost. In many cases we do not know who the observer was. In some cases it is noted in the chronicle that the record had been noted by the director of the Astronomical Office -- but this person does not necessarily have been the observer: It could have been any other astronomer at the court or anywhere in the empire. The contribution of an object to divine purposes only derives from its appearance at a certain place which might be a star or constellation, or an area as big as the circumpolar circle, or even one of the right ascension slices called lunar mansion. 

Aiming to evaluate the historical records to find classical novae and other transient phenomena among them, we need to check the accuracy of the positions given in these texts. 

\subsection{Position derived from historical record}
Out of the three collections of astronomical text passages from the Far Eastern chronicles (\cite{hsi}, \cite{ho}, and \cite{xu2000}) \citet{xu2000} is the most comprehensive and includes sources (chronicles) which had not been checked by Ho and Hsi. Thus, we rely only on this edition. From this collection, we have extracted the information on the position of the transient object. 

With this information, we derive the position in the sky as exact as possible and draw a circle in the same way as for the nova candidate search, i.\,e. either covering the asterism or -- in case of single stars -- an appropriate circle around them. 

 If the identified SNR or pulsar (according to Simbad data base) for the event lays within our defined circle, the angular distance would be zero. If not, we estimate the distance of the post-supernova object to the circumference or slightly within it: As the circles are selected areas with well reasoned radii but not considered as exact as measurements, the distance to the position in the text should be given only in full degrees (with no decimal places) and they are not defined as shortest distance to its rim. 

 \textbf{An example:} Consider the supernova in 185 CE. The text is given in an earlier paragraph and the position is found like explained above. The identified SNR G315.0-02.3 is next to $\alpha$ Cen but roughly 2\degr\ south of it. Thus, its position fits the (point) coordinates of Nanmen given by Stephenson (we assume a margin of error of 5\degr\ suggested by Stephenson's catalogue and plot a filled green circle in Fig.~\ref{fig:event185}). The problem is that none of the proposed combinations of two stars forming the asterism of Nanmen suggests a position south of $\alpha$ Cen. This can only be true if the position given in the text is not within the gate but next to the gate. Thus, the position of the SNR gives a margin of error of at least 1\degr\ or 2\degr\ (distance to our circle). 
 
 \textbf{Another example:} The \textbf{guest star in 1054} is the only one with a given position next to a single star and not within a constellation. The description in this case is preserved in six records cited in \citet[p.\,138/9]{xu2000}. Record (1) and (4) both say `several cun southeast of Tianguan' where `cun' is roughly a tenth of a degree. Records (2) and (3) give only the mantic description `guarding Tianguan' and records (5) and (6) describe it `in the space of' the lunar mansions Zui and Shen which are right ascension slices. For defining our circle we take the single star asterism Tianguan and as `several cun' should be less than one degree (because otherwise we suppose they would have said `1 du' equaling $0\fdg9856\approx1\degr$) we assume a radius of 1\degr . 
 
 The distance of M1 (Crab Nebula) and its pulsar to Tianguan ($\zeta$ Tau) is 1\degr 8\arcmin\ to the northwest. Hence, the direction differs from the description in the text by 180\degr\ but the distance roughly fits our circle (within the margin of error derived above). For the guest star in 1054 we derive an angular separation of the SNR from the position (circle) given in the text of roughly 1\degr\ (the circle's radius) because the text clearly says `southeast' instead of northwest. From the astrophysical point of view, \cite{mayall+oort} aiming to examine this discrepancy made a  search ``of the literature of Chinese uranographies, and the result is that the asterism T'ien kuan cannot be located much more precisely than `near $\zeta$~Tauri.' The idenity in position of the nova and nebula is therefore as close as a comparison of Oriental and Occidental uranographies will allow.'' \citep[p.\,101]{mayall+oort}. Additionally, such a mistake can also root in the copying tradition of the observer's original report: As the target is northwest of a reference star, the reference star is southeast of the target. Thus, an abbreviated note in a diary might be misleading and might have led to a misinterpretation by the chronicler who put it in a sentence in order to `explain' any political happening (and, therefore, preserved the observation). That means, only reading the text, we again would deduce a field in which the SNR is not located (southeast of $\zeta$ Tau) but taking into account the tradition of the text (searching in a full circle around $\zeta$ Tau and not only in one direction) and applying a margin of error of 1\degr\ or 2\degr\ the SNR position fits the description.
 
\subsection{Margin of error from comparison with modern counterparts}
Evaluating all eight supernova positions with our method, we define the circles to cover the area described in the text resulting in the list displayed in Tab.~\ref{tab:HIPradii}. We have chosen stars from the HIPPARCOS catalogue as centres of the circles because coordinates are shifted by precession even within the few decades between our predecessors (Hsi, 1957 to Stephenson$+$ in the 1970s) and us. By anchoring our suggestions with stars we hope to ease the comparison for later colleagues. The estimated angular separations are displayed in Tab.~\ref{tab:SNRangsep}.
\begin{table}
	\centering
	\caption{Coordinates of our derived circles for eight historical SNe (equinox 2000).}
	\label{tab:HIPradii}
	\fontsize{9}{10.8}\selectfont
	\begin{tabular}{rcrrrc} 
		\hline
		year & ID & HIP & RA & DE & radius/ \degr \\
		\hline
 185. &  & 68992. & 211.87 & -57.1284 & 6. \\
 386. & 1. & 92927. & 283.995 & -28.1302 & 3. \\
 386. & 2. & 90496. & 276.993 & -25.4217 & 3. \\
 386. & 3. & 92041. & 281.414 & -26.9908 & 2. \\
 386. & 4. & 89341. & 273.441 & -21.0588 & 3. \\
 393. &  & 84638. & 259.55 & -39.6908 & 5. \\
 1006. &  & 74376. & 227.984 & -48.7378 & 1. \\
 1054. &  & 26451. & 84.4112 & 21.1425 & 1. \\
 1181. &  & 7078. & 22.8073 & 70.2646 & 4. \\
 1572. & 1. & 3797. & 12.1975 & 60.7665 & 1. \\
 1572. & 2. & 4427. & 14.1772 & 60.7167 & 2. \\
 1604. &  & 84856. & 260.143 & -19.3337 & 4. \\
		\hline
	\end{tabular}
\end{table}

\begin{table}
	\centering
	\caption{Angular separations $\Delta\alpha$ of the post-supernova object (coordinates RA/DE here in equinox 20000) to the area of search which we derived from the position given in the text (Tab.~\ref{tab:tinynovacat}). $\Delta\alpha=0$ means, the remnant lays within our circles.}
	\label{tab:SNRangsep}
	\fontsize{9}{10.8}\selectfont
	\begin{tabular}{crlrrc} 
		\hline
		& year & SNR & RA & DE & $\Delta\alpha$/ \degr \\
		\hline
 1. & 185 & \text{SNR G315.0-02.3} & 220.75 & -62.5 &  2 \\
 2. & 386 & \text{SNR G011.2-01.1 } & 273.5 & -19.7667 &  0 \\
 3. & 393 & \text{SNR G347.3-00.5} & 258.113 & -39.6867  & 0 \\
 4. & 1006 & \text{Lupus SN} & 225.592 & -42.0969 & \textbf{6} (?) \\
 5. & 1054 & \text{PSR V* CM Tau} & 83.6331 & 22.0145 & 1 \\
 6. & 1181 & \text{SNR G130.7+03.1 } & 31.4043 & 64.8283  & 3 \\
 7. & 1572 & \text{SN1572A} & 6.3397 & 64.1408 & 4.5 \\
 8. & 1604 & \text{Kepler SNR} & 262.65 & -21.4823 & 0 \\
		\hline
	\end{tabular}
\end{table}

Concerning the supernova in 1006, we think the angular separation of the remnant from the position of about 6\degr\ in our table is misleading. This event was probably the brightest object seen by human beings since the dawn of written culture. It could easily have replaced any star in Lupus (not only $\kappa$ Lup). Although in modern time Qichen Jiangjun is the name for $\kappa$ Lup it might have been the name for another star in the epoch of the supernova. As the remnant is close to $\kappa$ Cen and $\beta$ Lup (two stars in the asterism of the Imperial Guards; the Cavalry General could be one of them), it suggests a wrong identification of the star mentioned in the text -- maybe even a modern exchange of $\kappa$ Cen and $\kappa$ Lup. The position described in the text fits the given degree in the lunar mansion and the directions to the other asterisms Di (around $\alpha$ Lib) and Kulou (around $\delta$ Cru, 29 stars). The asterism Qiguan (Cavalry Officers) is not clear: \citet[p.\,217]{ho} identifies it with the modern asterism of the Imperial Guards ($\gamma, \delta, \kappa$ Lup, $\beta$ Cen, $\lambda, \varepsilon, \mu , \pi, o, \alpha$ Lup). In this case, the Japanese record `entered Qiguan' would be correct, but the Chinese records `west of Qiguan' are not. \citet[p.\,423]{xu2000} identifies Qiguan with 27 stars around C1 Cen (some 3.5 hours west of Lupus) which certainly also does not fit. Thus, we resume that the whole description by asterisms is very uncertain and we will have to study the historical development of the maps in more detail than any previous astronomical study. For the moment, we just exclude this event from our analysis of the error bars of the positioning of events on the base of records. For the other events, the variances of descriptions and, therefore, the degrees of freedom for the interpretation are fewer.  

 The map of SN~185 has already been shown in \ref{fig:event185}. For the remaining seven supernovae we provide an appendix of their maps, containing the analysis with our procedure, as well as the suggestions by previous authors. The point coordinates given by \citet{stephenson} and \citet{pskovskii} are likely rounded by 5\degr\ because their numbers in declination always end with 0 or 5. Hence, we attach a circle of 5\degr\ to all of their point coordinates. As the numbers in the catalogue by \citet{xi+po} do not show such a pattern, we consider only 1\degr-circles around their point coordinates.

\section{Results and Conclusions}
Mapping these fields and the positions suggested by the authors of earlier lists of historical nova positions (Fig.~\ref{fig:overview}), we find that in many cases the earlier localizations differ from each other. This reflects their redundant trial to describe the areas given in the text by point coordinates (for pointing a modern telescope). Instead we argue, that the historical texts provide only area positions instead of definite points.  

From three recent collections of historical reports, we selected a list of 25 events (in 24 years selected from a text corpus ranging from $-600$ to $+1690$) with a rather high probability to be a nova because it is reported as star-like (no tail, no movement reported) and lasted for more than one night. In many cases of the roughly 600 original records, no duration is given: These events are neglected in our shortlist but have to be considered for a complete catalogue of possible novae. However, our short list is sufficient to develop an algorithm for how to proceed: We try to locate the position in the sky, i.\,e. define fields for a catalogue search for transient objects and variables. The result is a list of these fields (Tab.~\ref{tab:tinynovacat}). Some of the areas given in the reports are huge (several ten or a few hundred square degrees): Considering the fact that most of the historical texts do not preserve a position with regard to a single star but only with regard to a constellation measuring some ten to some hundred square degrees, the fields in which we are looking for the modern counterpart become huge. 

 The margin of error for these fields turned out (Tab.~\ref{tab:SNRangsep}) to be 4\fdg5 from the angular separation of the remnants of the identified cases of historical supernovae (neglecting the SN\,1006 value because of the wrong identification of the star name). 

  For example, consider the cases of a guest star in the asterism of Beidou (equaling the `Big Dipper'), e.\,g. the guest star in 329 CE (as well as 158 and 305 CE which are not considered in our shortlist): In these cases the text forces us to survey an area of roughly 10\degr $\times$ 30\degr\ (the area of the Dipper). To this huge field, we have to apply a margin of error of (at least) 4\fdg5 in each direction as derived from the known historical supernovae obtaining an even bigger field of 19\degr $\times$ 39\degr. In several cases the applied error bar might lead to an area extending to (or even including) a neighboring asterism. This has to be checked by plotting the whole scenario (asterism, all found CVs in the given area, lines of neighboring asterisms) in a map. 
 
 However, the fields in which we search remain impractically huge for telescopic observations. Hence, we suggest to focus observations on certain stars in the field: Presuming classical novae we would search for cataclysmic variables already discovered within these areas as well as X-ray binaries and possibly mistaken planetary nebulae. Yet, as we could never be sure about the cause of an appearance the areas should also be checked for flare stars and any other type of variables. Thus, our next goal would be a list of all possible remnants of classical novae (such as CVs, X-ray binaries, PN) within these fields and a further goal should be a list of other types of outbursting stars we could imagine. Withal, as catalogues are continuously updated it is not worth to print any list of objects but only the areas in the sky in which the event is reported to have taken place. We would like to point out that this resulting list of fields in which the historical events have happened is a very good approach but not the final answer: This has to be done for all historical records in all lists including the records which are currently considered to describe comets in order to possibly find atmospherically blurred bright appearances. This has to be technically linked to continuous catalogue requests to propose targets for successful follow-up observations.   

\section{Outlook}
 A database with positions of transients in the historical records is going to be published online. It will contain the cases presented in this paper as well as all the other 118 transients in the record collection of \citet[p.\,129--146]{xu2000} which are supposed to have not been comets. However, in none of the cases, the historical report provides any description of the development of the phenomenon to be sure that it really was a classical nova. It could also have been a flare star, a rarely observed Mira-like, or any other type of transient -- in some cases even a tailless comet (although this is highly unlikely due to our selection criteria but in some cases not completely excludable). 

 As the areas given in historical text are expansive, it is not practical and hardly possible to derive specific observational targets only from the historical text. We can only derive the area in the sky where the transient was seen at the time and leave it to further research to decide whether the historical transient had been a classical nova, a supernova, or maybe none of this but a comet, a meteor, a flare star or some other phenomenon. That means, all objects found in catalogues of nova and supernova remnants might lead to a wrong conclusion concerning the physics of close binaries as long as we cannot exclude the possibility that the object was a tailless comet or any type of variable star or transient phenomenon. In some cases this might be highly unlikely but it is not decidable with only these records. Hence, the research strategy should be:
\begin{enumerate} 
 \item define the fields (areas!) in which we are looking for modern counterparts of the appearances, 
 \item check (regularly) all catalogues of CVs, X-ray binaries, Mira-likes, flare stars, supernova-remnants, nova-shells, planetary nebulae (which could be classified incorrectly), etc., 
 \item answer the question which type of object the historical record really reports, 
 \item if and only if this can be stated securely: derive information on the physics of the object or even the physics of a certain type of objects.
\end{enumerate} 
  
\section*{Acknowledgements}
S.H. thanks the Free State of Thuringa for financing the project at the Friedrich Schiller University of Jena, Germany. N.V. acknowledges financial support from FONDECYT regular No. 1170566 and from Centro de Astrofísica, Universidad de Valparaíso, Chile. The authors thank Jesse Chapman (Department of East Asian Languages and Cultures, University of California, Berkeley) for consulting us in terms of ancient Chinese astral science and religion. Thankfully we made use of the VSX variable star catalogue of the American Association of Variable star Observers (AAVSO) and of the SIMBAD data base (Strassbourg). Ralph Neuh\"auser (AIU, Friedrich-Schiller-Universität Jena) had the initiative and idea to reconsider historical nova identifications including new nova candidates in a transdisciplinary project. We thank him to have brought us together.

\subsection*{Author contributions}
This analysis of historical texts and search for modern counterparts of possible classical novae has been performed by SH and NV together. PP contributed the maps in Fig.~3 and supported the considerations of naked eye observations with his master thesis on the reconstruction of methods to observe the ancient star catalogue. 

\bibliography{chinGuestStars_2}%

\appendix
\newpage
Additionally, we append all other seven maps of historical positions estimated by four authors (Stephenson$+$, Pskovskii, Hsi, Xi and Po; green, purple, yellow, gray respectively) of the last few decades in comparison with our own estimated circles (green line). Above the upper rim of the map the starting year of the appearance is written and -- if given -- its duration (in brackets). The asterism(s) mentioned in the text (right) are marked in the map and written on the right rim. The red dot indicates the position of the associated SNR. 

\begin{figure}
	\includegraphics[width=\columnwidth]{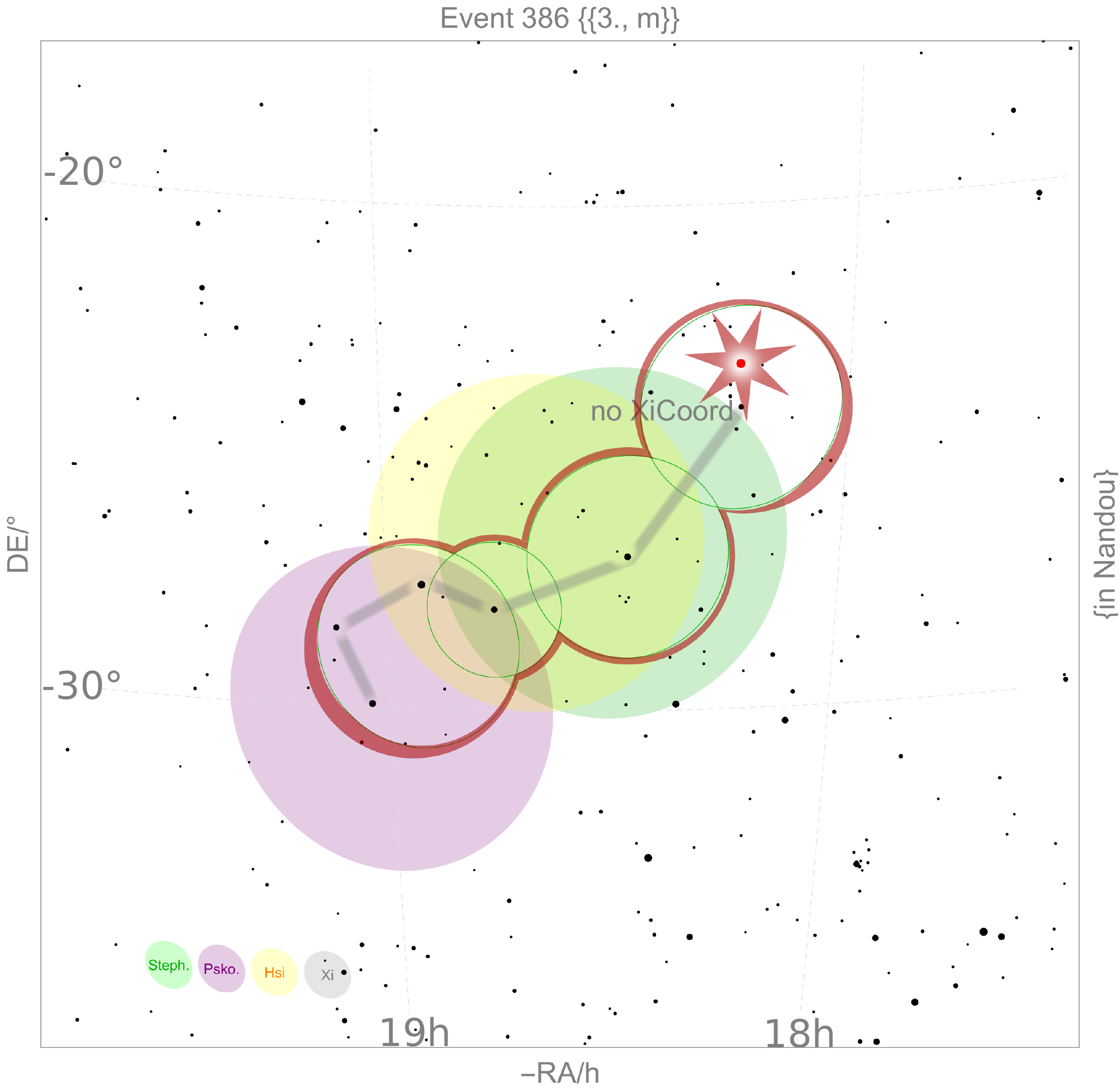}
    \caption{Map of the positions interpreted by the considered authors: The textual description `There was a guest star in Nandou.' is excerpted at the right rim. In brackets after the year is the duration (3 months) if given; in this case: `3rd month. (\dots) until the 6th month'. The position of the SNR is marked with a red dot and highlighted by a star with seven spikes.}
    \label{fig:event386}
\end{figure}

\begin{figure}
	\includegraphics[width=\columnwidth]{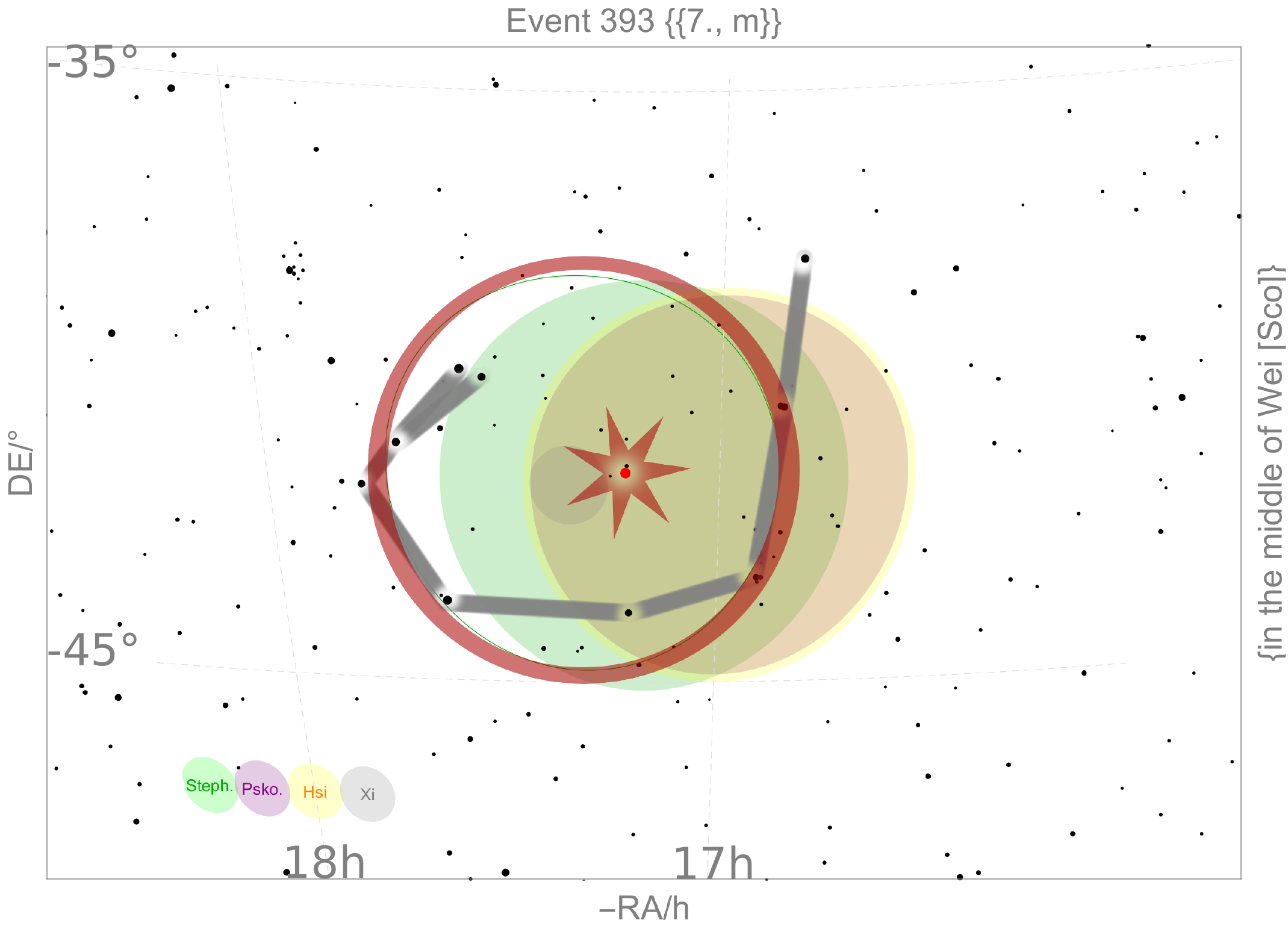}
    \caption{Map of the positions interpreted by the considered authors. The text gives `in the middle of Wei' and `2nd month (\dots) until 9th month'. The position of the SNR is marked with a red dot and highlighted by a star with seven spikes.}
    \label{fig:event393}
\end{figure}

\begin{figure}
	\includegraphics[width=\columnwidth]{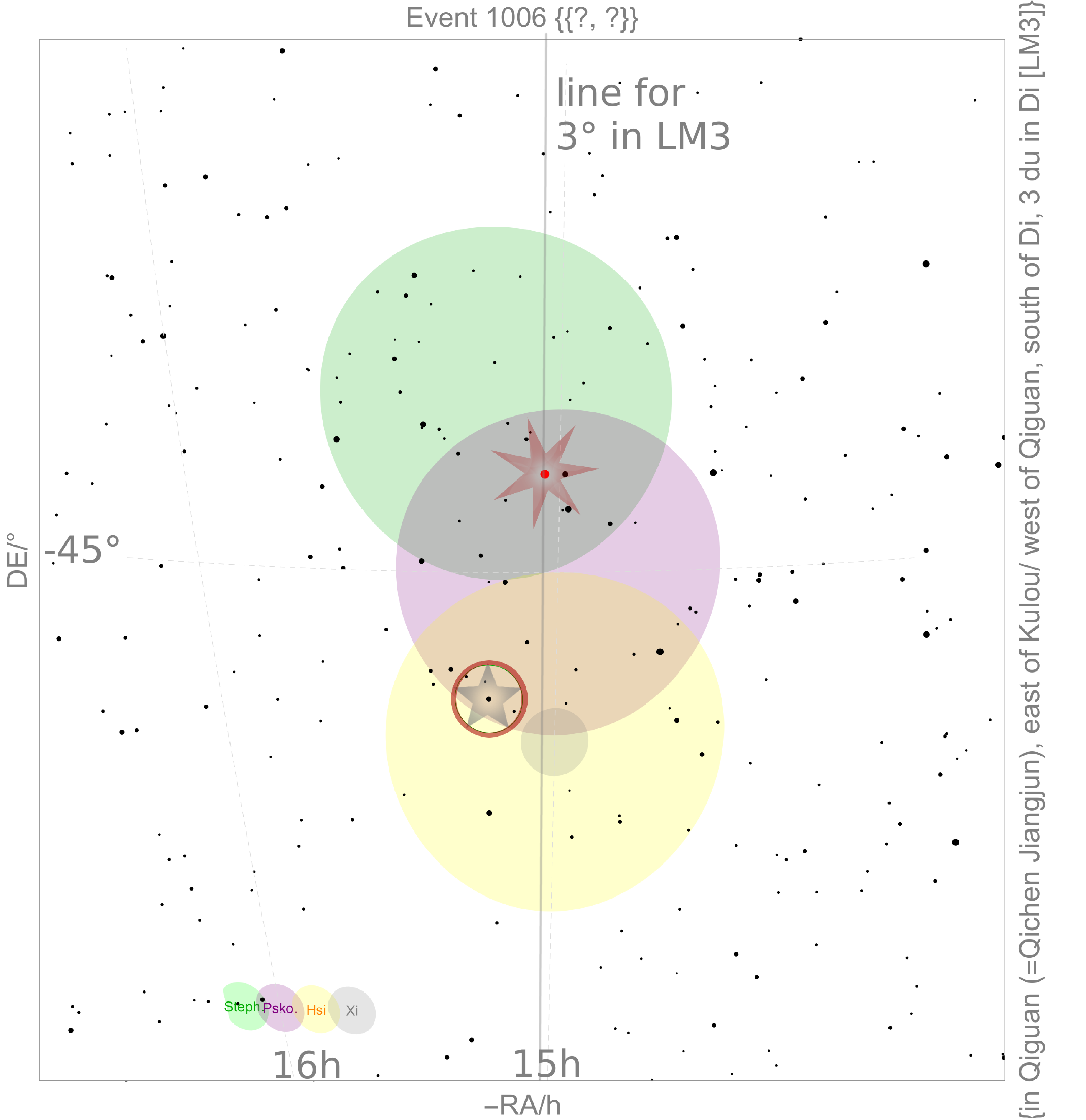}
    \caption{There are three descriptions by constellations and one measurement of a right ascension (cf. right edge of figure) but no mentioning of a disappearance or duration. Qichen Jiangjun (The Cavalry General) is usually identified with at least $\kappa$ Lupi. It can be a single star asterism or an assembly of this star with two others close by (according to Rufus (1945), p.\,17) `one of the brighter stars of Lupus', according to Ho (1962), p.\,217 $\kappa$ Lup, according to Xu et al. (2000), p.\,423 three stars around $\kappa$ Lup). Kulou's main star is $\delta$ Cru (to the west of this map, consistent). The lunar mansion Di is defined by $\alpha$ Lib. Hence, the asterisms Di is far north of the upper rim of this map (consistent). The measured RA fits the SNR while the various descriptions by asterisms are confusing. The position of the SNR is marked with a red dot and highlighted by a star with seven spikes.}
    \label{fig:event1006}
\end{figure}

\begin{figure}
	\includegraphics[width=\columnwidth]{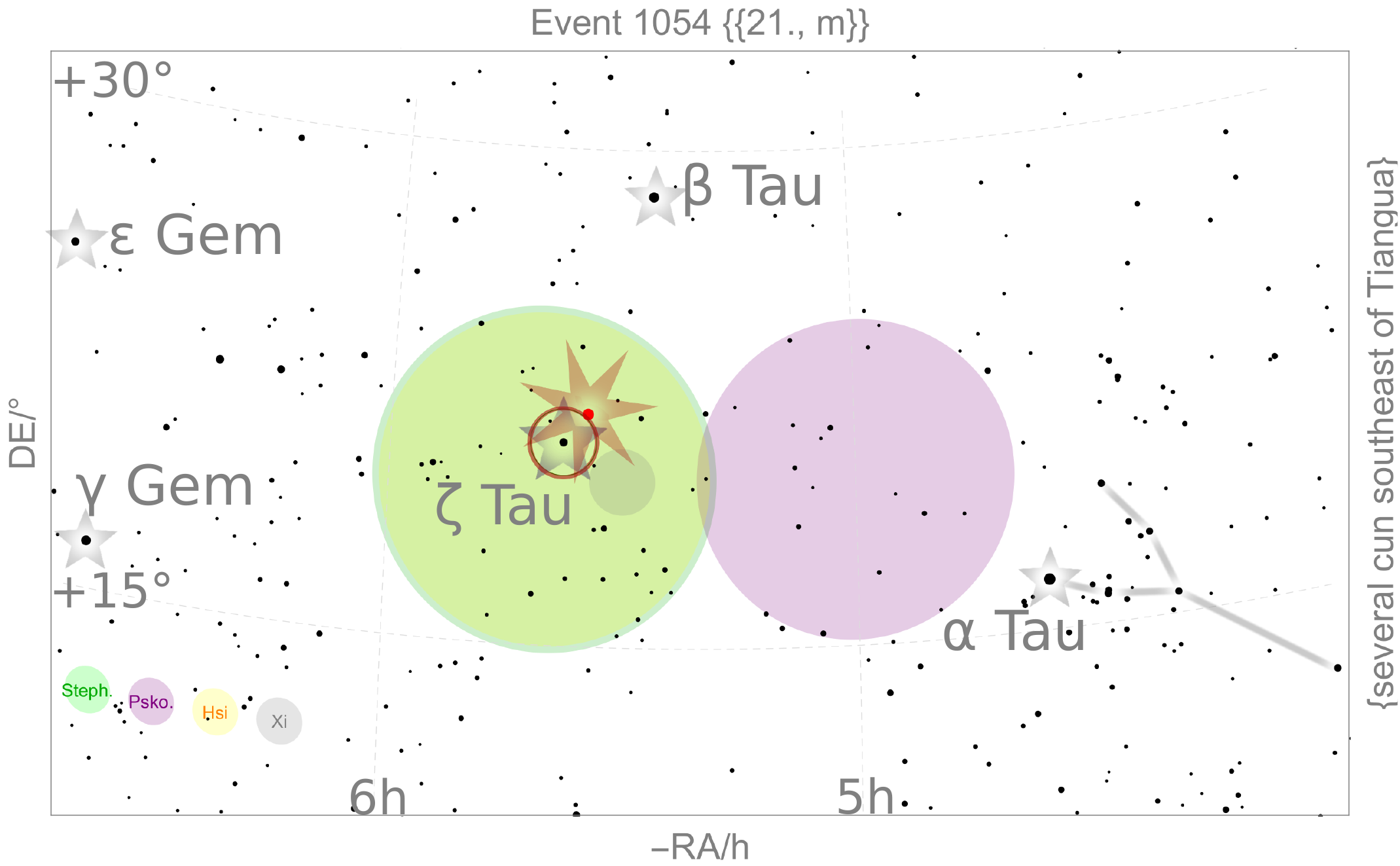}
    \caption{There are six records preserved (according to Xu et al. (2000), p.\,138--139), the most precise position comes from China `A guest star emerged several cun southeast of Tianguan.' The duration of 21 months is derived from the date of disappearance. Tianguan (The Celestial Pass) is a single star asterism ($\zeta$ Tau). For orientation please note the Hyades at the right edge of the map. The position of the SNR is marked with a red dot and highlighted by a star with seven spikes.}
    \label{fig:event1054}
\end{figure}

\begin{figure}
	\includegraphics[width=\columnwidth]{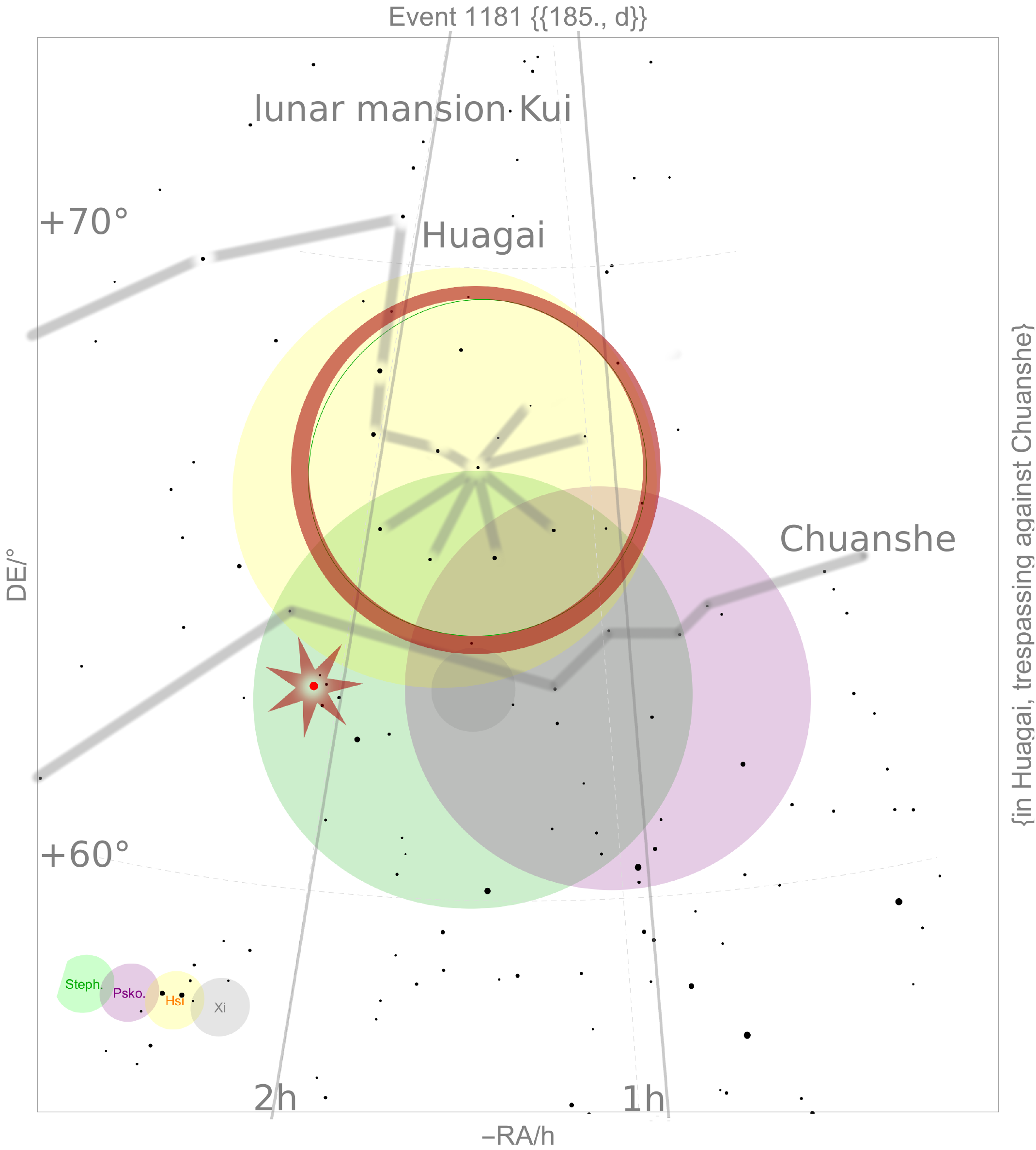}
    \caption{Two texts from China give `in Huagai' and `in lunar mansion Kui and trespassing against Chuanshe' lasting 185 (or 156) days; a record from Japan describes it `close to Wangliang guarding Chuanshe' (note that Wangliang asterism is close to Chuanshe, including the three bright stars of Cassiopeia in the lower right corner of this map). Trespassing against an asterism means approaching it to within one du. The position of the SNR is marked with a red dot and highlighted by a star with seven spikes.}
    \label{fig:event1181}
\end{figure}

\begin{figure}
	\includegraphics[width=\columnwidth]{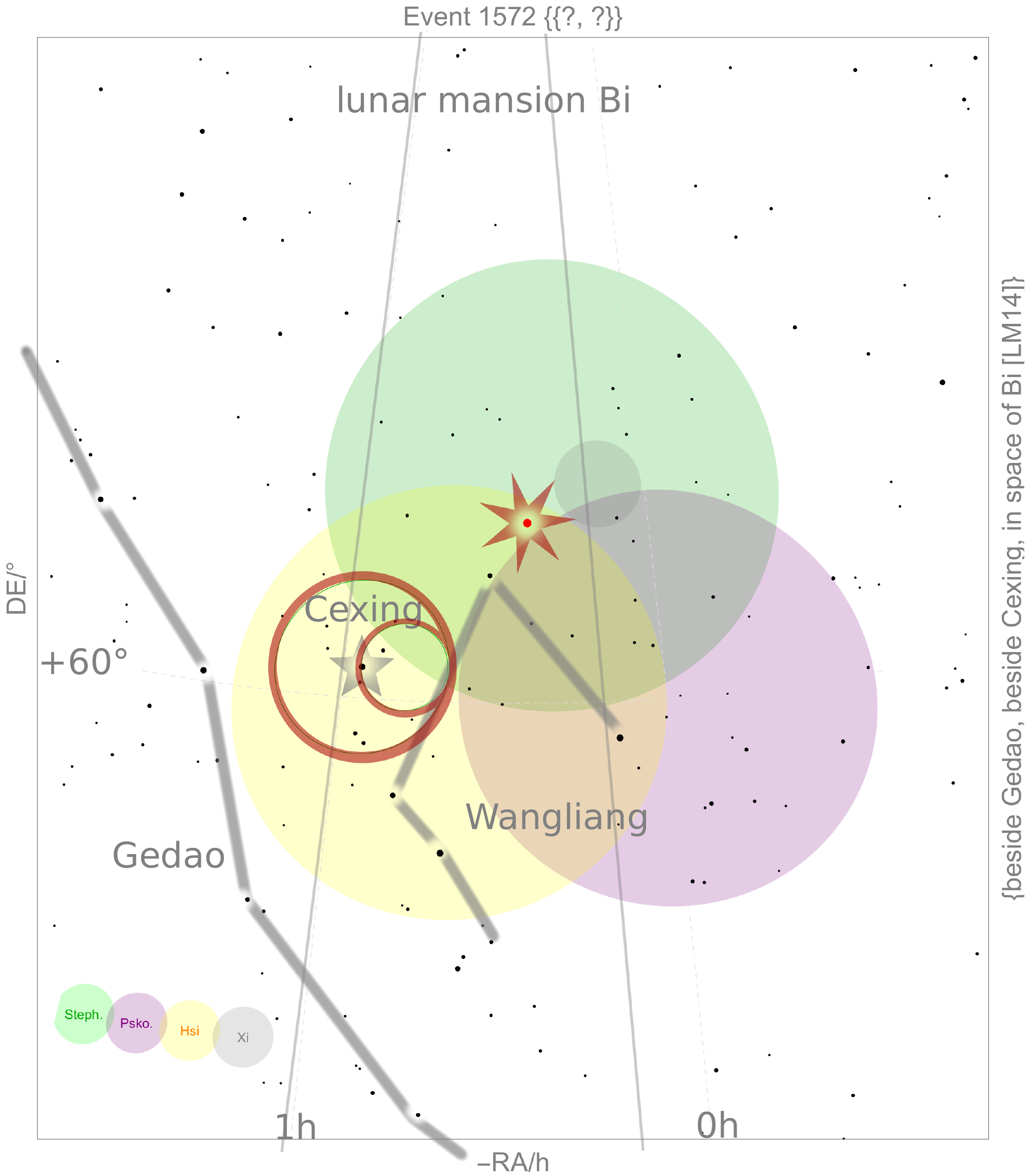}
    \caption{One record from China describes a `guest star (\dots) beside Gedao in space of Bi.' A record from Korea reports it `beside Cexing' which is a single star asterism ($\gamma$ Cas). A duration of the appearance is not given. The position of the SNR is marked with a red dot and highlighted by a star with seven spikes.}
    \label{fig:event1572}
\end{figure}

\begin{figure}
	\includegraphics[width=\columnwidth]{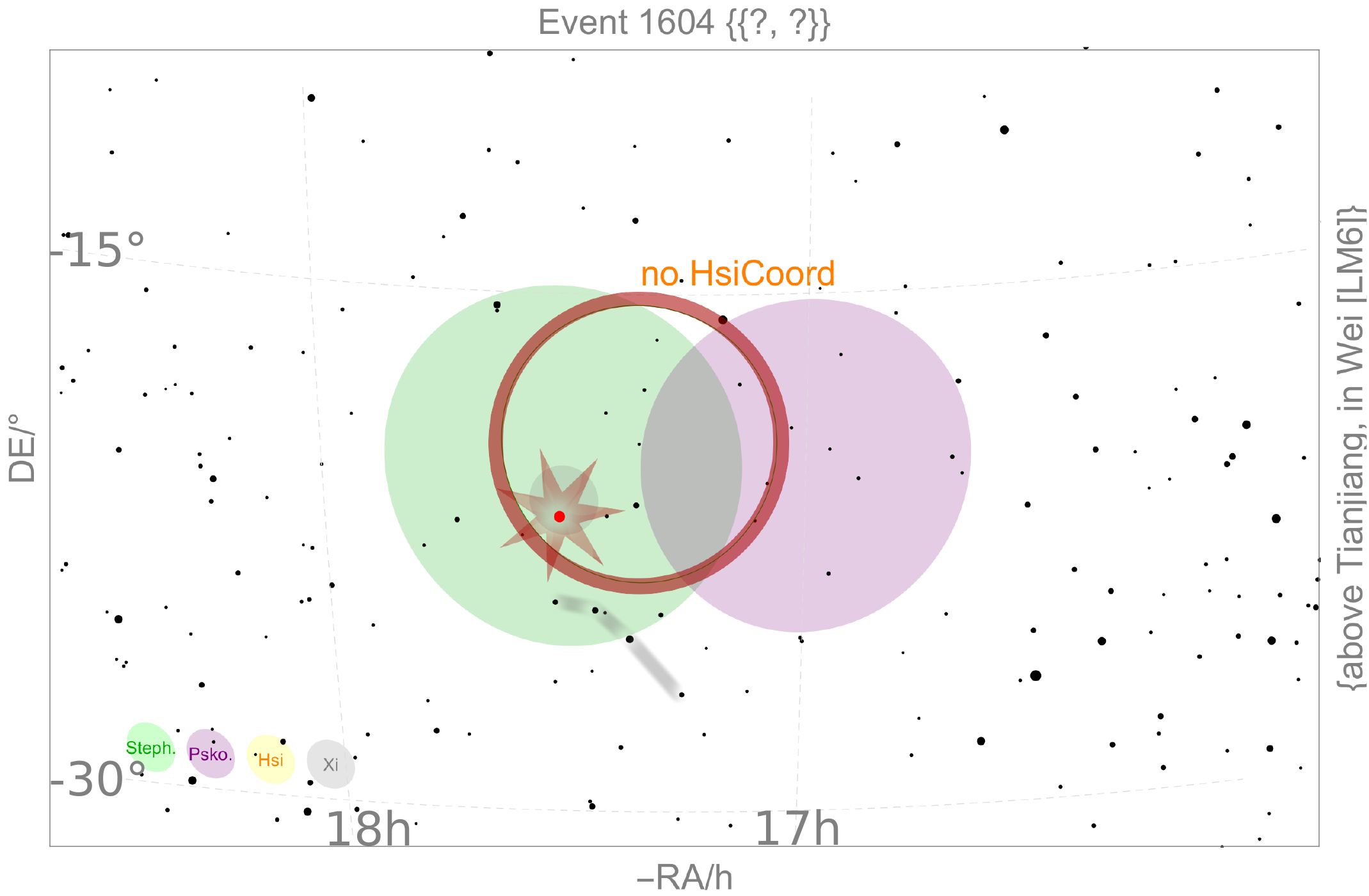}
   \caption{Four records from China, two from Korea (according to Xu et al. (2000)) describe a guest star `in Wei' (probably the lunar mansion) and `above Tianjiang'. The small asterism of Tianjiang contains b, 36 and $\theta$ Oph. The position of the SNR is marked with a red dot and highlighted by a star with seven spikes.}
    \label{fig:event1604}
\end{figure}

\end{document}